\documentclass[a4paper,11pt]{article}
\usepackage{jheppub} 
\usepackage{bbm,bm,graphicx,mathtools,color,slashed}
\usepackage{amssymb,amsmath}
\usepackage{hyperref}
\usepackage[T1]{fontenc}

\hypersetup{
setpagesize=false,
 bookmarksnumbered=true,%
 bookmarksopen=true,%
 colorlinks=true
}

\usepackage[all]{xy}

\usepackage[framemethod=tikz]{mdframed}
\usepackage{tikz}
\usepackage{tikz-feynhand}

\usepackage[normalem]{ulem}  

\newcommand{\tr}{\mathop{\mathrm{tr}}}
\newcommand{\Tr}{\mathop{\mathrm{Tr}}}
\newcommand{\im}{\mathop{\mathrm{Im}}}

\newcommand{\diff}{\mathrm{d}} 
\newcommand{\rmi}{\mathrm{i}} 
\newcommand{\rme}{\mathrm{e}}

\newcommand{\etas}{\lambda_s}

\newcommand{\ha}{\mathrm{a}}
\newcommand{\hb}{\mathrm{b}}
\newcommand{\hc}{\mathrm{c}}

\newcommand{\ds}{\displaystyle}

\newcommand\Lcal{\mathcal{L}}

\newcommand\Ocal{\mathcal{O}}

\newcommand\Ncal{\mathcal{N}}

\newcommand{\average}[1]{\langle#1\rangle}

\renewcommand\Re{\mathop{\mathrm{Re}}}

\newcommand\SU{{\rm SU}}

\newcommand{\bx}{\bm{x}}

\newcommand{\bk}{\bm{k}}
\newcommand{\bp}{\bm{p}}
\newcommand{\bq}{\bm{q}}

\newcommand{\br}{\bm{r}}
\newcommand{\bv}{\bm{v}}
\newcommand{\bs}{\bm{s}}
\newcommand{\bD}{\bm{D}}
\newcommand{\bJ}{\bm{J}}
\newcommand{\bnab}{\bm{\nabla}}
\newcommand{\bzero}{\bm{0}}

\newcommand{\with}{\quad\mathrm{with}\quad}

\newcommand{\be}{\begin{equation}}
\newcommand{\ee}{\end{equation}}

\title{
Spin relaxation rate for heavy quarks in weakly coupled QCD plasma
}

\author[a,b]{Masaru Hongo,}
\author[c,d]{Xu-Guang Huang,}
\author[e]{Matthias Kaminski,}
\author[a]{Mikhail Stephanov,}
\author[a]{Ho-Ung Yee}

\affiliation[a]{Department of Physics, University of Illinois, Chicago, IL 60607, USA}
\affiliation[b]{RIKEN iTHEMS, RIKEN, Wako 351-0198, Japan}
\affiliation[c]{Physics Department and Center for Field Theory and Particle Physics, Fudan University, Shanghai 200433, China}
\affiliation[d]{Key Laboratory of Nuclear Physics and Ion-beam Application (MOE), Fudan University, Shanghai 200433, China}

\affiliation[e]{Department of Physics and Astronomy, University of Alabama, Tuscaloosa, AL 35487, USA}

\emailAdd{hongo@uic.edu, huangxuguang@fudan.edu.cn, mski@ua.edu, misha@uic.edu, hyee@uic.edu
}

\abstract{
We compute the relaxation rate of the spin density of heavy quarks in a perturbative QCD plasma to leading-log order in the coupling constant $g$. 
The spin relaxation rate $\Gamma_s$ in spin hydrodynamics is shown to be $\Gamma_s\sim g^4\log(1/g)T (T/M)^2$ in the heavy-quark limit $T/M\ll 1$, which is smaller than the relaxation rate of other non-hydrodynamic modes by additional powers of $T/M$.
We demonstrate three different methods to evaluate the spin relaxation rate: 1) the Green-Kubo formula in the spin hydrodynamic regime, 2) the spin density correlation function in the strict hydrodynamic limit, and 3) quantum kinetic theory of the spin distribution function in momentum space. 
We highlight the interesting differences between these methods, while they are ultimately connected to each other by the underlying Ward-Takahashi identity for the non-conserved spin density.
}

\begin{document}
\maketitle

\section{Introduction} 
\label{sec:intro}

The recent experimental observation of the spin polarization of hadrons in the rotating QCD plasma produced in off-central relativistic heavy-ion collisions at RHIC and LHC~\cite{STAR:2017ckg,STAR:2019erd,ALICE:2019aid,STAR:2020xbm} has motivated much theoretical and phenomenological study of the dynamical evolution of spin polarization in a finite temperature plasma. 
A large amount of theoretical effort has been invested into the development of a description of spin in kinetic theory~\cite{
Gao:2019znl,Weickgenannt:2019dks,Hattori:2019ahi,Wang:2019moi,Li:2019qkf,Kapusta:2019sad,Liu:2019krs,Yang:2020hri,Liu:2020flb,Weickgenannt:2020aaf,Weickgenannt:2021cuo,Sheng:2021kfc,Lin:2021mvw} and spin in hydrodynamics~\cite{Florkowski:2017ruc,Florkowski:2018fap,Hattori:2019lfp,Fukushima:2020ucl,Bhadury:2020puc,Shi:2020htn,Li:2020eon,Gallegos:2021bzp,Liu:2021uhn,She:2021lhe,Hongo:2021ona,Peng:2021ago}. 
As the spin angular momentum is not conserved and is transferable to orbital angular momentum, the spin density, in general, shows a relaxation behavior toward its local equilibrium value in hydrodynamics~\cite{Hattori:2019lfp,Hongo:2021ona}. 
The local equilibrium value of spin density is dictated by thermodynamics with rotation, i.e., with conserved total angular momentum~ \cite{Becattini:2007nd,Becattini:2009wh,Becattini:2014yxa,Palermo:2021hlf}.  
An interesting aspect of this equilibrium state is that the chemical potential corresponding to conserved total angular momentum is equal to the local thermal vorticity of fluids~\cite{Becattini:2007nd,Becattini:2009wh}.
In the strict hydrodynamic regime, the local spin polarization is determined by the local thermal vorticity via a thermodynamic relation. 
Moreover, spin hydrodynamics in this regime is shown to be equivalent, by a pseudo-gauge transformation~\cite{Becattini:2012pp,Becattini:2018duy, Florkowski:2018fap, Speranza:2020ilk}, to conventional hydrodynamics with a symmetric energy-momentum tensor with a set of non-dissipative transport coefficients involving fluid vorticity~\cite{Li:2020eon,Fukushima:2020ucl}. 

The relaxation rate of the spin density toward its equilibrium value is an important quantity in determining how the spin polarization evolves in time in theoretical simulations of QCD plasma.
The QCD plasma produced in heavy-ion collisions goes through several different phases in its lifetime, and one would need to study the dynamics of spin in all phases to reliably predict the observed value for the spin polarization of hadrons. In this work, we study the spin relaxation rate in the quark-gluon plasma phase, where we assume a high enough temperature to apply a weakly coupled description of QCD plasma, i.e., the finite temperature field theory of perturbative QCD (pQCD)~\cite{Braaten:1989mz,LeBellac2000,Blaizot:2001nr}.
Although this is not a realistic assumption for the plasma in heavy-ion collisions
where the coupling constant may not be small, the result provides a valuable benchmark in one extreme limit of the theory. 

We will focus on the relaxation of spins carried by heavy quarks in the limit $M\gg T$, where $M$ is the heavy-quark mass, and $T$ is the temperature of the plasma.
There are two simplifications in this case. 1) The density of heavy quarks in the plasma is dilute, and we may neglect interactions between heavy quarks. The relaxation of heavy quark spin results from its interactions with other background thermal particles. 
As we will see, the dominant contribution to the leading-log result comes from scatterings with light hard particles of momentum of order $T$. 
2) The spin relaxation rate in this limit will be shown to be of the order 
$\Gamma_s\sim g^4\log(1/g) T \left( T/M \right)^2$, where $g$ is the QCD coupling constant. It is smaller than the relaxation rates $\Gamma$ of other non-hydrodynamic modes by additional powers of $T/M\ll 1$.
This is an important fact that allows us to introduce the spin hydrodynamics in the regime of frequency scale $\Gamma_s\ll \omega \ll \Gamma$, where the spin density appears as an additional independent quasi-hydrodynamic degree of freedom with a relaxation behavior: an example of the Hydro+ description~\cite{Stephanov:2017ghc}. 
The relaxation of spin density in this regime of spin hydrodynamics is governed by a new kinetic coefficient $\etas$~\footnote{We note that $\etas$ is the heavy-quark (i.e., non-relativistic) counterpart of the rotational viscosity $\eta_s$ introduced in relativistic spin hydrodynamics~\cite{Hattori:2019lfp,Fukushima:2020ucl,She:2021lhe,Gallegos:2021bzp,Hongo:2021ona}. However, different from $\eta_s$ which is a transport coefficient in the relativistic case, the $\etas$ [see eq.~(\ref{eq:sa-gamma}) for its definition] in the non-relativistic case does not represent any transport phenomena but is merely an Onsager kinetic coefficient as we will explain in section \ref{sec:formulation}.}, which appears in the constitutive relations of spin hydrodynamics.

The suppression of spin relaxation rate in the heavy-quark limit can be understood in the non-relativistic limit of heavy-quark dynamics since, in the hydrodynamic limit of a near thermal equilibrium state, the typical heavy-quark velocity is small, $v=p/M\sim \sqrt{T/M}\ll~1$. In the heavy-quark limit, the QCD Lagrangian involving the heavy quark and gluon is reduced to
\begin{equation}
 \Lcal 
  = \rmi \psi^\dag D_0 \psi 
  - \frac{1}{2M}(\bm D \psi)^\dag \cdot \bm D \psi 
  + \frac{g}{2M} \psi^\dag (\bm B\cdot \bm\sigma)\psi 
  + \Lcal_{\mathrm{gluon}}
  + {\cal O}(1/M^2), 
  \label{eq:action}
\end{equation}
where $\psi$ is the non-relativistic two-component spinor, $D_\mu \psi= \partial_\mu \psi - \rmi g A_\mu \psi$ with gluon field $A_\mu$, $B^\ha=\epsilon^{\ha\hb\hc}F_{\hb\hc}/2$ ($\ha=1,2,3$ and $\epsilon^{\ha\hb\hc}$ is the Levi-Civita symbol) is the color magnetic field with the field strength tensor $F_{\mu\nu} \equiv \partial_\mu A_\nu - \partial_\nu A_\mu - \rmi g [A_\mu,A_\nu]$, $\sigma^\ha$ is the Pauli matrix, $\Lcal_{\mathrm{gluon}}=-(1/2)\tr F^{\mu\nu} F_{\mu\nu}$ is the gluon Lagrangian, and a term $-M \psi^\dag \psi$ is not shown. 
The first two leading terms, which are counted as of order $T$, conserve the non-relativistic $\SU(2)$ heavy-quark spin symmetry~(see, e.g., ref.~\cite{Manohar-Wise2000}), and the leading spin-violating interaction is given by the third term --- the Pauli term responsible for the coupling between spin and color magnetic fields.
The vital point for our discussion is that the Pauli term is suppressed by the additional power of $T/M\ll 1$,%
\footnote{
The thermal fluctuations of the gluon field $A_\mu$ are at most of the order $T$, which can also be seen by a naive dimensional counting that the energy dimension of $A_\mu$ is 1. 
This gives $\bm B$ fluctuations at most of the order $T^2$.
}
and this term gives rise to the spin relaxation rate. 
For our computation of the spin relaxation rate to leading order of the $1/M$ expansion, it is enough to work with this non-relativistic action, which is simpler than the original relativistic theory of quarks as discussed in ref.~\cite{Li:2019qkf}. 
We will henceforth take this non-relativistic theory as the starting point in our computation of the spin relaxation rate in pQCD. Note that the thermal field theory of the gluon field $A_\mu$, as well as other light species of quarks, is still relativistic.

There have been a few previous works on the spin relaxation rate in finite temperature pQCD, based on the kinetic-theory framework~\cite{Li:2019qkf,Kapusta:2019sad,Liu:2019krs,Yang:2020hri, Weickgenannt:2020aaf,Weickgenannt:2021cuo,Sheng:2021kfc,Lin:2021mvw}
(see also ref.~\cite{Hidaka:2022dmn} for a recent review of quantum kinetic theory). 
These works describe the dynamics of the spin of quarks distributed in phase space as a result of scatterings with other thermal particles. 
Our objective in this work is to compute the kinetic coefficient responsible for the relaxation of spin density, i.e., the heavy-quark rotational viscosity $\etas$, in the macroscopic description of spin hydrodynamics in the heavy-quark limit $M\gg T$. 
Although it is in principle determined by the microscopic collision term obtained in the previous works,
the result has not been available in the literature.

We present three different methods in pQCD to compute the spin relaxation rate of heavy quarks in spin hydrodynamics, $\Gamma_s={\etas/\chi_s}$, where $\chi_s$ is the spin susceptibility.  
It is assuring that all three methods give the same result, but the reason why they all should agree is not at all obvious, at least superficially. It is ultimately a consequence of the underlying Ward-Takahashi identity for spin density, together with the separation of scales between $\Gamma_s$ and other relaxation rates $\Gamma$, which allows us a subtle controlled limit of frequency $\omega$ in the perturbative evaluation of spin density correlation function. 

The organization of the paper is as follows.
In section~\ref{sec:formulation}, we give a brief review of the retarded spin density correlation function
predicted by spin hydrodynamics. 
We discuss the Ward-Takahashi identity for the non-relativistic heavy-quark spin symmetry, which allows us to identify the quantum operator responsible for the spin relaxation in the macroscopic description. 
Based on this, in section~\ref{description}, we describe our three different methods of computing the spin relaxation rate $\Gamma_s$, in the leading-log order of the QCD coupling constant.
In section~\ref{sec:computation}, we present the details of computation in these methods and show that they all lead to the same result.
Section~\ref{sec:summary} is devoted to our summary and discussions.
In appendix \ref{sec:Kadanoff-Baym}, we present a derivation of the quantum kinetic equation for heavy quarks based on the Kadanoff-Baym formalism.

\section{Correlation functions in spin hydrodynamics} 
\label{sec:formulation}

Our purpose is to consider the dynamics of spin density attached to the non-relativistic fermion $\psi$.
The crucial point here is that the Lagrangian \eqref{eq:action} 
enjoys the approximate heavy quark $\SU(2)$ symmetry that acts on $\psi$ as 
$\psi \to \rme^{\rmi \theta^{\ha} \sigma_{\ha}} \psi$. Only the Pauli (third) term in eq.~\eqref{eq:action} breaks such heavy quark symmetry, and one can show the following Ward-Takahashi identity as the equation of motion for the spin density operator:
\begin{equation}
 \partial_0 J^0_{\ha} + \bnab \cdot \bJ_{\ha}
  = \Theta_{\ha} \quad 
  (\ha = 1,2,3),
  \label{eq:spin-non-conservation}
\end{equation}
where we introduced the spin current $J^\mu_{\ha}$ and the source term as 
\begin{equation}
 J^\mu_{\ha} = 
  \begin{pmatrix}
   \frac{1}{2}\psi^\dag \sigma_{\ha} \psi \\
   - \frac{\rmi}{4M} 
   \big[ 
   \psi^\dag \sigma_{\ha} (\bD \psi)  - (\bD \psi)^\dag \sigma_{\ha} \psi 
   \big]
  \end{pmatrix}, \quad 
   \Theta_{\ha} \equiv 
   - \frac{g}{2M} \epsilon_{\ha\hb\hc}  \psi^\dag B^{\hb}\sigma^{\hc} \psi.
\end{equation}
Equation \eqref{eq:spin-non-conservation} indicates that 
spin for the heavy quark is not conserved due to the Pauli term between the spin and color magnetic field.
However, it is crucial to note that the source term $\Theta_{\ha}$ has the factor $1/M$, so that it is suppressed in the heavy-quark limit.
In other words, one can regard eq.~\eqref{eq:spin-non-conservation} as an approximate conservation law if $M$ is large enough.

We can coarse-grain operator equation~(\ref{eq:spin-non-conservation})
to obtain the relationship between the coarse-grained variables for
which we shall use the same notations as for the operators. 
The existence of hydrodynamic description implies that all coarse-grained
operators can be expressed as local functionals of the conserved
densities such as $J^0_{\ha}$ as well as $\Theta_{\ha}$ using constitutive equations. 
The form of these equations is strongly constrained by the second law of
thermodynamics, which requires that the entropy functional $S[J^0_\ha]=\int \diff^3 x s(J^0_\ha)$ is not decreasing in time. 
Using coarse-grained eq.~(\ref{eq:spin-non-conservation}), we can express 
the local entropy production rate as 
\begin{equation}
  \label{eq:TdS}
  \partial_0 s 
  + \bnab \cdot \bs 
  = 
   \bJ_{\ha} \cdot \bnab \frac{\partial s}{\partial J^0_{\ha}}
  + \Theta_{\ha} 
  \frac{\partial s}{\partial J^0_{\ha}}
  \ge 0
  \with 
  \bs \equiv \frac{\partial s}{\partial J^0_{\ha}}
  \bJ_{\ha} ,
\end{equation}
for which we required the local second law of thermodynamics.
This means that both the source $\Theta_{\ha}$ and current $J^i_a$ must be proportional to 
\begin{equation}
 \label{eq:dsdJ}
  T\frac{\partial s}{\partial J^0_{\ha}} \equiv -\mu_{\ha} + b_{\ha},
\end{equation}
and its spatial gradients, where we defined internal spin chemical potential $\mu_{\ha}$, which is a
function of spin density $J^0_{\ha}$, as well as external potential for
the spin $b_{\ha}$, which could be the fluid thermal vorticity, or the external magnetic field times gyromagnetic
ratio, or even the torsion. We note that the external spin potential $b_{\ha}$ is different from the dynamical magnetic field $B_{\ha}$. The second law of thermodynamics in eq.~(\ref{eq:TdS}) thus
constrains the constitutive equations to leading order in gradients up
to semi-positive-definite kinetic coefficients:
\begin{subequations}\label{eq:constit}
 \begin{equation}
  \label{eq:Jia-sigma}
    \bJ_{\ha} = - 
    T \sigma_s 
    \bnab 
    \left( \frac{\mu_{\ha} - b_{\ha} }{T}\right),
 \end{equation}
 \begin{equation}
  \label{eq:sa-gamma}
   \Theta_{\ha} = - 
   \etas \left( \mu_{\ha} - b_{\ha} \right).
 \end{equation}
\end{subequations}
Substituting constitutive equations~(\ref{eq:constit}) into the approximate conservation law (\ref{eq:spin-non-conservation}), 
we obtain hydrodynamic equation of motion for spin density $J^0_a$:
\begin{equation}
 \partial_0 J^0_{\ha}
  - \bnab \cdot
  \left[ 
  T \sigma_s 
  \bnab \left( \frac{\mu_{\ha} - b_{\ha}}{T} \right)
  \right]
  = - \etas \left( \mu_{\ha} - b_{\ha} \right).
  \label{eq:spin-eom}
\end{equation}

Equilibrium is achieved at a value of $J^0_{\ha}$ at which the internal
and external spin potentials are equal, i.e., $\mu_{\ha} = b_{\ha}$. 
At this point, the entropy is maximized, according to eq.~(\ref{eq:dsdJ}) and
does not increase, according to eq.~(\ref{eq:TdS}). 
Linearizing equation of motion~(\ref{eq:spin-eom}) in small deviations around that equilibrium, we find
\begin{equation}
  \label{eq:J0a-linearized}
  \partial_0 \delta J^0_{\ha}
  - D_s \bnab^2 \delta J^0_{\ha} 
  + \sigma_s \bnab^2 \delta b_{\ha} =
  - \Gamma_s \delta J^0_{\ha} + \etas \delta b_{\ha} ,
\end{equation}
where we introduced a spin diffusion coefficient $D_s$ and spin relaxation rate $\Gamma_s$ as 
\begin{equation}
 D_s\equiv \frac{\sigma_s}{\chi_s}
  \quad \mbox{and } \quad
  \Gamma_s \equiv  \frac{\etas}{\chi_s}
  \label{eq:Ds,Gs}
\end{equation}
using the spin susceptibility $\chi_s$  defined as usual by
\begin{equation}
 \frac{\partial \mu_{\ha}}{\partial J^0_{\hb}}
  = \delta_{\ha\hb} \,\chi_s^{-1} .\label{eq:chis}
\end{equation}

Equation~(\ref{eq:J0a-linearized}) describes linear response of the spin
density to perturbations of the external spin potential, and we can
use it to determine the retarded Green's function for spin density~\cite{Kadanoff-Martin1963}:
\begin{equation}
 \label{eq:GR-Jb}
   G^{J^0_{\ha} J^0_{\hb}}_R (\omega,\bm k)
  = \frac{\rmi (\etas + \sigma_s \bm k^2 )}{\omega + \rmi
  (\Gamma_s + D_s \bm k^2 )} \delta_{\ha\hb} .
\end{equation}
We then find that the retarded spin density correlator has a pole at the imaginary value of 
\begin{equation}
 \omega (\bm k) = - \rmi ( \Gamma_s + D_s \bm k^2 ) + O (\bm k^4),
\end{equation}
which does not vanish in $\bm k \to \bzero$ limit.  Therefore, in
hydrodynamic limit $\bm k\to \bzero$ the spin density shows relaxational
behavior with characteristic time $\tau_s = \Gamma_s^{-1}$. 

If the spin relaxation time $\tau_s$ is much longer than other microscopic time scales, then the hydrodynamic regime $\omega\ll \Gamma$ can be split into two subregimes. 
The strict hydrodynamic regime, i.e., $\omega$ is much
smaller than any relaxation scales, including $\omega\ll\Gamma_s$, and
the regime where $\Gamma_s\ll\omega\ll \Gamma$. 
The latter is the so-called
Hydro$+$ regime~\cite{Stephanov:2017ghc}, where a small subset of non-hydrodynamic modes
relaxes on a scale comparable to the hydrodynamic time scale. 
Correspondingly, there are two ways to obtain the spin relaxation rate $\Gamma_s = \etas/\chi_s$ as we shall explain below.

By taking a strict hydrodynamic limit $\omega\to 0$ at $\bm k = \bzero$ of the spin density
correlator~\eqref{eq:GR-Jb} and using eq.~\eqref{eq:Ds,Gs}, we obtain one way to evaluate the spin relaxation rate as~\cite{Hongo:2021ona}
\begin{equation}
 \lim_{\omega \to 0}
  \im \frac{1}{\omega} G_R^{J^0_{\ha} J^0_{\hb}} (\omega, \bm k =\bzero) 
  = \frac{\chi_s}{\Gamma_s} \delta_{\ha\hb}. 
  \label{eq:Green-Kubo1}
\end{equation}
Alternatively, we can take the limit $\omega\ll \Gamma$, while still
maintaining $\omega \gg \Gamma_s$:
\begin{equation}
 \lim_{\Gamma_s\ll\omega\ll \Gamma}
  \im {\omega} G_R^{J^0_{\ha} J^0_{\hb}} (\omega, \bm k =\bzero) 
  = {\etas} \delta_{\ha\hb}. 
  \label{eq:ImGR+}
\end{equation}
Since the imaginary part of the retarded Green's function is related
to the fluctuation correlator $G_{rr}$ through the fluctuation-dissipation relation $G_{rr}= (2T/\omega) \im G_R$ , 
we can rewrite both eqs.~(\ref{eq:Green-Kubo1}) and~(\ref{eq:ImGR+}) in
terms of $G_{rr}^{J^0_{\ha} J^0_{\hb}}$.
Moreover, since, at $\bm k=\bzero$, by eq.~(\ref{eq:spin-non-conservation}),
$\Theta_{\ha} = \partial_0 J^0_{\ha}$, the correlator of the spin density $J^0_{\ha}$ can be related to the correlator of the source $\Theta_{\ha}$ by a Ward-Takahashi identity, and we obtain
\begin{equation}
 \label{eq:JJ-SS}
  G_{rr}^{\Theta_{\ha} \Theta_{\hb}} 
  (\omega, \bm k =\bzero) =
  \omega^2 G_{rr}^{J^0_{\ha} J^0_{\hb}} (\omega, \bm k =\bzero)\,.
\end{equation}
Using this identity and the fluctuations-dissipation relation we can rewrite
eq.~(\ref{eq:ImGR+}) in terms of the symmetric correlator of $\Theta_{\ha}$:
\begin{equation}
 \label{eq:gammas-GrrSS}
 \frac{1}{2T}
 \lim_{\Gamma_s\ll\omega\ll \Gamma}
  G_{rr}^{\Theta_{\ha} \Theta_{\hb}} 
  (\omega, \bm k =\bzero ) 
  = \etas \delta_{\ha\hb},
\end{equation}
which gives another useful formula to evaluate heavy-quark rotational viscosity $\etas$.

\section{Description of the methods}
\label{description}

Before we present our detailed computations of the spin relaxation rate in three different methods in section~\ref{sec:computation}, let us give a brief overview of these methods, which summarizes the main ideas, as well as the differences between them.

Based on the previous section, let us start with the retarded Green's function of spin density in zero wavenumber limit $\bm k\to \bzero$:
\be
G_R^{J^0_{\ha} J^0_{\hb}} (\omega)
  = \frac{\rmi \chi_s \Gamma_s }{\omega + \rmi \Gamma_s } \delta_{\ha\hb},
  \label{eqint1}
\ee
where we use the shorthand notation for the zero wave number Green's function as $G (\omega) = G (\omega,\bk=\bzero)$.
The same correlation function computed in finite temperature pQCD must agree with (\ref{eqint1}) in both spin and strict hydrodynamic regimes in $\omega\ll \Gamma \sim g^4\log(1/g)T$. 
The fact that $\Gamma_s\sim \Gamma \left( \frac{T}{M} \right)^2 \ll \Gamma$ 
allows us to consider two different regimes of $\omega$, i.e., 
the spin hydrodynamic one $\omega \gg \Gamma_s$ and 
the strict hydrodynamic one $\omega\ll \Gamma_s$. 

\paragraph{(1) The Green-Kubo formula in spin hydrodynamic regime:}
In the spin hydrodynamic regime, we have an expansion in powers of $\Gamma_s/\omega$ as
\be
G_R^{J^0_{\ha} J^0_{\hb}} (\omega)
  = \left( 
  \frac{\rmi \chi_s \Gamma_s }{\omega} 
  + {\chi_s\Gamma_s^2\over\omega^2} 
  + \cdots
  \right)
  \delta^{\ha\hb}
  \quad \mathrm{at} \quad
  \Gamma_s \ll \omega \ll \Gamma.
  \label{eqint2}
\ee
The first term can also be regarded as the result of taking the controlled limit we discussed in the previous section, i.e., $\lim_{\Gamma_s\ll\omega\ll \Gamma}G_R^{J^0_{\ha} J^0_{\hb}}(\omega)$.
We see that the series is organized in terms of increasing powers of $\Gamma_s$, which is equivalent to increasing powers of the coupling constant $g$.
Since $\Gamma_s\to 0$ in the perturbative limit $g\to 0$, the condition for the expansion to work, i.e., $\omega\gg \Gamma_s$, is valid for any fixed non-zero value of $\omega$. This means that the diagrammatic perturbation series of the correlation function in field theory should be one-to-one correspondent to the expansion in eq.~(\ref{eqint2}), as long as we keep $\omega$ fixed and finite.
Especially, matching the first term in eq.~(\ref{eqint2}) with the first leading diagrams in naive perturbation theory in $g$, we are able to compute $\Gamma_s$ in simple perturbative computations.
We emphasize that this is possible only because the spin density is not a conserved quantity, and we have a finite non-vanishing relaxation rate $\Gamma_s$ that is perturbative in $g$.
This is in sharp contrast to the correlation functions for conserved quantities, for which the density-density correlation function in $\bm k= \bzero$ limit simply vanishes identically for all $\omega$ due to the Ward-Takahashi identity. 
 
In the diagrammatic computation, it turns out to be easier to compute the Wightman correlation function $G^{J^0_{\ha} J^0_{\hb}}_{12}(\omega)$, which is related to the retarded correlation function $G_{R}^{J^0_{\ha} J^0_{\hb}}(\omega)$ by the fluctuation-dissipation relation in $\omega\ll T$ regime,
\be
 G_{12}^{J^0_{\ha} J^0_{\hb}} (\omega) 
 = {2T\over \omega} \im G^{J^0_{\ha} J^0_{\hb}}_R (\omega),
\ee
where $T/\omega$ comes from the limit of Bose-Einstein distribution $n_B(\omega)$ in the small frequency limit $\omega\ll T$.
Therefore, our discussion above implies that the leading diagram for $G_{12}^{J^0_{\ha} J^0_{\hb}}$ in the naive perturbation theory should match to $2T\chi_s\Gamma_s\delta_{\ha\hb}/ \omega^2$ in $\Gamma_s \ll \omega\ll T$ regime.
As we discussed in the previous section, one can, in fact, use the Ward-Takahashi identity to relate the correlation functions of spin density $J^{0}_{\ha}$ with those of source operator $\Theta_{\ha}$, i.e., $\partial_0 J^{0}_{\ha}= \Theta_{\ha}$ in $\bm k=\bzero$ limit. 
In frequency space, this implies $\omega^2 G^{J^0_{\ha} J^0_{\hb}}_{12}(\omega)=G^{\Theta_{\ha} \Theta_{\hb}}_{12}(\omega)$, which we have also checked diagrammatically for our leading order diagrams. 
The former involves three diagrams, while the latter turns out to be only one diagram in leading order. 

We, therefore, use the latter to compute the spin relaxation rate, by matching 
\begin{equation}
  G_{12}^{\Theta_{\ha} \Theta_{\hb}}(\omega \to 0) = 2T\chi_s\Gamma_s\delta_{\ha\hb},
  \label{eq:12-Gamma-spni}
\end{equation}
with the leading order diagram in the naive perturbation theory in $g$.

Note that this is the Wightman expression of eq.~\eqref{eq:gammas-GrrSS}.
As we show in the next section in detail, we are able to determine $\Gamma_s$ in leading-log of coupling constant based on this formula as
\be
 \Gamma_s 
   = C_2 (R) \frac{g^2 m_D^2 T}{6 \pi M^2} \log (1/g),
   \label{eqint3}
\ee
where $m_D^2={g^2 T^2}(2N_c+N_F)/6$ is the Debye mass squared, and $C_2(R)$ is the Casimir invariant of the color representation $R$ of the heavy quark, which is $C_2(F)=(N_c^2-1)/(2N_c)$ for fundamental representation of $\SU(N_c)$.

\paragraph{(2) The spin density correlation function in the strict hydrodynamic regime: }

On the other hand, in the strict hydrodynamic regime of $\omega\ll \Gamma_s$, 
which is the true hydrodynamic limit of $\omega\to 0$, 
we have an expansion in $\omega/ \Gamma_s$ as
\be
 G_R^{J^0_{\ha} J^0_{\hb}} (\omega) 
 = \left(
 \chi_s + \rmi{\chi_s\over\Gamma_s} \omega + {\cal O}(\omega^2)
 \right)\delta_{\ha\hb}
 \quad \mathrm{at} \quad
 \omega \ll \Gamma_s,
 \label{eq:true-hydro-spin-spin}
\ee
and the same $\Gamma_s$ can be computed from the leading imaginary part in small frequency limit.
In this limit, the imaginary part of the retarded correlation function becomes sensitive to infrared singularities arising from the diverging mean-free path of heavy quarks in $g\to 0$ limit, which is called the pinching singularities~\cite{Jeon:1994if}.

The pinching singularity should be regulated by including the imaginary part of  self-energy in the heavy-quark propagators, i.e., the damping rate, which represents the scatterings with thermal background medium that gives rise to a finite mean-free path. 
The proper evaluation of correlation functions in this limit further requires a resummation of infinite ladder diagrams~\cite{Jeon:1994if,ValleBasagoiti:2002ir,Jimenez-Alba:2015bia}, i.e., the vertex corrections, in addition to the damping rate in the propagators: only after this, the Ward-Takahashi identity (or the conservation law in the case of exact global symmetry) is fulfilled~\cite{Aarts:2002tn,Hidaka:2010gh}. 
One can also understand the pinching singularity from the Boltzmann equation.
In the language of kinetic theory, the damping rate and the vertex correction correspond to the loss and the gain terms, respectively, in the collision terms of the Boltzmann equation.
The appearance of $\Gamma_s$ in the denominator, which brings in a non-analyticity of the correlation function in coupling $g$, is a manifestation of pinching singularities of spin density correlation function that is regulated by a finite relaxation rate of spin density. 

The fact that only $\Gamma_s$, not $\Gamma$ in general, appears in the spin density correlation function after resuming an infinite number of ladder diagrams is also an interesting difference from the case of conserved charges, where the density-density correlation function at $\bk =0$ is simply zero. 
The case of conserved charges can be understood by replacing $\Gamma_s$ with an infinitesimal $\epsilon$ that goes to zero. 
In this limit, the imaginary part linear in $\omega$ seems to have a divergent coefficient ${1/\epsilon}$, but the condition of expansion $\omega\ll \epsilon$ is never justified. The proper thing to do is to go back to eq.~(\ref{eqint1}) with $\Gamma_s$ replaced by $\epsilon$, and the density-density correlation function vanishes in $\epsilon\to 0$ limit. 
The transport coefficients for conserved charges, e.g., shear viscosity and conductivity, are computed from the current-current correlation functions instead, which shows the similar non-analytic behavior of $1/\Gamma\sim 1/[g^4 T\log(1/g)]$ from resummation of infinite ladder diagrams. The Ward-Takahashi identity for
spin density should be responsible for how the spin density correlation function has a similar but different non-analytic behavior of $1/\Gamma_s$ after resummation, which depends only on the spin-violating interactions that are suppressed by powers of $T/M$.

Another way to look at the difference between $\omega\gg \Gamma_s$ and $\omega\ll \Gamma_s$
in the diagrammatic evaluation of correlation functions is the following. 
In the expansion in eq.~(\ref{eqint2}), which is organized by the naive perturbation theory in $g$, each term in the series becomes of the same order when $\omega\sim \Gamma_s$, and one needs to include all terms to properly evaluate the correlation function in $\omega\ll \Gamma_s$ limit. Diagrammatically, this necessitates a summation over an infinite number of diagrams. The relevant subset of diagrams in leading order is captured by the diagrams with pinching singularities in the reorganized perturbation theory with damping rate included in the propagators.

\paragraph{(3) The quantum kinetic theory for spin: }

Our final method of computing the spin relaxation rate of heavy quarks is the quantum kinetic theory of spin density matrix, developed in ref.~\cite{Li:2019qkf}. 
The time evolution of a spin distribution function in momentum space, $\bm S(\bm p, t)$, is described by the quantum Boltzmann equation with collision terms,
\be
{\partial \bm S(\bm p,t)\over\partial t}=\hat{\bm \Gamma}_S[\bm S(\bm p,t)]
\ee
where $\hat{\bm \Gamma}_S$ is the quantum collision operator acting on the spin distribution function (not to be confused with $\Gamma_s$). The leading-log expression for $\hat{\bm\Gamma}_S$ is available for a massive quark with its mass satisfying the condition $M\gg gT$, and we can take $M\gg T$ limit for our purpose. The result is organized in powers of $T/M$, 
\be
 \hat{\bm\Gamma}_S=\hat{\bm\Gamma}_S^{(0)}+ \hat{\bm\Gamma}_S^{(1)}+\cdots ,
 \label{eq:quantum-kinetic}
\ee
and the leading term reproduces the momentum diffusion equation with the heavy-quark drag force known in literature~\cite{Moore:2004tg}.

The leading term conserves the total spin density, i.e., 
$\int_{\bm p} \hat{\bm\Gamma}_S^{(0)}[\bm S(\bm p)]=0$ with $\int_{\bm p}\equiv\int \diff^3 p/(2\pi)^3$ for any $\bm S(\bm p)$, and the spin relaxation in leading order is given by the next term $\hat{\bm\Gamma}_S^{(1)}$. The eigenmodes of $\hat{\bm\Gamma}_S^{(0)}$ have eigenvalues of order $\sim g^4\log(1/g)T(T/M)$, which are non-hydrodynamic modes of the microscopic theory, except the zero mode, $\bm S^{(0)}(\bm p)=\bm S_0 \rme^{-\beta E_p}$ with $E_p={\bm p}^2/(2M)$ for any constant vector $\bm S_0$, that has zero eigenvalue of $\hat{\bm\Gamma}_S^{(0)}$. The zero mode represents the equilibrium distribution of spin polarization in momentum space in leading order, which takes a simple Boltzmann distribution. In other words, the zero mode coefficients $\bm S_0$ corresponds to the spin density in the spin hydrodynamics. How the zero mode relaxes by the spin violating term, $\hat{\bm\Gamma}_S^{(1)}$, gives us the spin relaxation rate in hydrodynamics, $\Gamma_s$.
Writing $\bm S(\bm p,t)=\bm S_0(t) \rme^{-\beta E_p}$, and inserting this leading order expression to the quantum kinetic equation~\eqref{eq:quantum-kinetic}, we obtain, to the leading order in $1/M$, 
\be
 {\partial \bm S_0(t)\over\partial t} = - \Gamma_s \bm S_0(t)
 \with
 \Gamma_s=- {\int_{\bm p}  \hat{\Gamma}_S^{(1)}[ \rme^{-\beta E_p}]\over
 \int_{\bm p} \rme^{-\beta E_p}}.
\ee
We used the fact that $ \hat{\bm\Gamma}_S^{(1)}[\bm S_0 \rme^{-\beta E_p}]= \bm S_0\hat{\Gamma}_S^{(1)}[ \rme^{-\beta E_p}]$ with a scalar operator  $\hat{\Gamma}_S^{(1)}$, due to rotational invariance of the collision term.
We emphasize that the sub-leading corrections in $1/M$ to the expression, $\bm S(\bm p,t)=\bm S_0(t) \rme^{-\beta E_p}$, exists in general, but are removed upon integration over $\bm p$, due to the fact that $\int_{\bm p} \hat{\bm\Gamma}_S^{(0)}[\bm S(\bm p)]=0$ for any $\bm S(\bm p)$. This is the procedure that identifies $\Gamma_s$ correctly to leading order in $1/M$ in quantum kinetic theory.

\section{Evaluation of the spin relaxation rate in perturbative QCD} 
\label{sec:computation}

In this section, 
we present our detailed evaluations of the spin relaxation rate in three different ways, all of which lead to the same result given in~\eqref{eqint3}.
We first compute $\Gamma_s$ via the Green-Kubo formula 
of the source correlation and the spin density correlation functions in two different regimes, $\omega\gg \Gamma_s$ and $\omega\ll\Gamma_s$, in sections~\ref{sec:Green-Kubo} and~\ref{sec:spin-spin}, respectively.
In section~\ref{sec:kinetic}, we evaluate the spin relaxation rate in yet another way, based on the quantum kinetic theory of the spin distribution function in momentum space.

\subsection{Method 1: Green-Kubo formula in $\omega\gg\Gamma_s$ regime}
\label{sec:Green-Kubo}

As we described in the previous section, the spin relaxation rate can be computed by the source-source correlation function, $G_{12}^{\Theta_{\ha} \Theta_{\hb}} (\omega\to0)$, evaluated in the naive perturbation theory, and matching that with $2T\chi_s \Gamma_s\delta_{\ha\hb}$, where $\chi_s$ is the spin susceptibility [recall eq.~\eqref{eq:12-Gamma-spni}].

Let us first summarize the basic ingredients for diagrammatic computation.
The real-time propagators for the non-relativistic heavy quarks and the relativistic gluons are 
\begin{equation}
 \begin{split}
  S_{12} (k)
  &= 
  \parbox{2.5cm}{\vspace{8pt}\includegraphics[width=2.5cm]{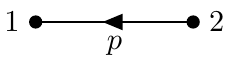}}
  = - n_F (k^0) 2 \pi \delta (k^0 - E_{\bk}),
  \\
  S_{21} (k)
  &= 
  \parbox{2.5cm}{\vspace{8pt}\includegraphics[width=2.5cm]{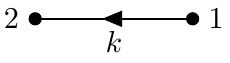}}
  = \big[ 1 - n_F(k^0) \big]
  2 \pi \delta (k^0 - E_{\bk}),
  \\
  D^{\mu\nu}_{12} (k)
  &= 
  \parbox{2.5cm}{\vspace{12pt}\includegraphics[width=2.5cm]{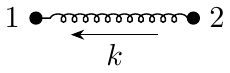}}
  = n_B (k^0) \rho^{\mu\nu} (k) ,
  \\
  D^{\mu\nu}_{21} (k)
  &=
  \parbox{2.5cm}{\vspace{12pt}\includegraphics[width=2.5cm]{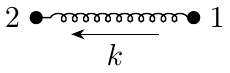}}
  = \big[ 1+ n_B (k^0) \big] \rho^{\mu\nu} (k) ,
  \end{split}
  \label{eq:prop}
\end{equation}
where $E_{\bk} = \bk^2/2M$ is the non-relativistic kinetic energy of heavy quark, $\rho^{\mu\nu}(k)$ is the gluon spectral density, and  
the Fermi-Dirac and Bose-Einstein distribution functions are
$n_F (k^0) = 1/(\rme^{\beta k^0} \rme^{\beta M} + 1)$ and 
$n_B (k^0) = 1/(\rme^{\beta k^0} - 1)$, respectively. 
For simplicity, we use the notation $k^\mu=(k^0,\bk)$ (thus, $k^0$ corresponds to $\omega$ in previous sections), and omit the trivial color structures.
Note that the appearance of fugacity $\rme^{-\beta M}$ in $n_F$ is due to the fact that the non-relativistic energy is measured from the rest mass energy $M$.%
\footnote{We assume that the (relativistic) heavy-quark chemical potential is zero, which leads to a dilute limit $n_F\sim \rme^{-\beta M}\ll 1$ when $T \ll M$.}
 
First of all, we note that the source operator $\Theta_{\ha}=-(g/ 2M) \epsilon_{\ha\hb\hc} \psi^\dagger B^{\hb} \sigma^{\hc} \psi$ already contains the small parameter $g/M$, and it also carries the gluon field. 
The leading diagram in the naive perturbation theory is the 2-loop sunset diagram [see figure in eq.~\eqref{eq:gamma-integral} below].
Since the gluon field in $\Theta_{\ha}$ carries a non-zero frequency-momentum in general (of order $gT$ as we will see), the two propagators of heavy quarks do not carry the same frequency-momentum, and there is no infrared singularity of pinching poles in the computation. 
Therefore, we can safely use the free propagator for heavy quarks in our leading order evaluation. 

Moreover, due to the on-shell conditions imposed on the heavy-quark momenta, as can be seen in eqs.~(\ref{eq:prop}), the gluon momentum is constrained to be spacelike. The free gluon spectral density has support only on light-like momenta.
The leading correction to the gluon spectral density $\rho^{\mu\nu}(k)$ for spacelike momenta comes from the Hard Thermal Loop (HTL) self-energy, which is well-known in literature~\cite{Braaten:1989mz,LeBellac2000,Blaizot:2001nr}. 
Our computation is based on these ingredients. 

The simple scattering picture is also possible by cutting the sun-set diagram in half.
The resulting diagram simply represents the $t$-channel scattering of an on-shell heavy quark with hard thermal particles of momenta $T$ in the background plasma. It turns out that the leading-log comes from the soft $t$-channel scatterings, with exchanged gluons carrying momenta in the range $m_D\ll |\bm k| \ll T$. 
This is similar to other leading-log results for transport coefficients~\cite{Baym:1990uj,Arnold:2000dr}.

For convenience, we here summarize the HTL results~\cite{Braaten:1989mz} (see, e.g., refs.~\cite{LeBellac2000,Blaizot:2001nr} for reviews on the HTL). 
The HTL spectral density for gluons 
takes the following form,
\begin{equation}
 \rho^{\mu\nu} =  
  \big[ 
  \delta^{\mu 0} \delta^{\nu 0}  \rho_L (k)
  + P^{\mu\nu}_T (\bm k) \rho_T (k)
  \big]
  \with
  P^{\mu\nu}_T(\bm k) \equiv 
  \delta^{\mu 0} \delta^{\nu 0} + \eta^{\mu\nu} - \frac{\bm k^\mu \bm k^\nu}{\bk^2}. 
  \label{eq:spectral-gluon-decomposition}
\end{equation}
where the longitudinal and transverse spectral functions are 
\begin{equation}
  \rho_L (k) 
  = - 2 \im \left[\frac{1}{\bk^2 - \Pi_L (k)}\right]
  \quad  \mathrm{and} \quad
  \rho_T(k) 
  = 2 \im \left[\frac{1}{\bk^2 - (k^0)^2 - \Pi_T (k)}\right], \quad 
\end{equation}
with the corresponding HTL self energies given by 
\begin{equation}
 \begin{split}
  \Pi_L(k)
  &= - m_D^2 
  \left[ 
   1 + \frac{k^0}{2|\bk|} \log \frac{k^0 - |\bk| + \rmi \epsilon}{k^0+|\bk| + \rmi \epsilon} 
  \right],
  \\
  \Pi_T(k) 
  &= - \frac{m_D^2}{2} 
  \left[ 
  \frac{(k^0)^2}{|\bk|^2}
  + \left( \frac{(k^0)^2}{|\bk|^2} - 1 \right)
  \frac{k^0}{2|\bk|}
  \log \frac{k^0 - |\bk| + \rmi \epsilon}{k^0+|\bk| + \rmi \epsilon} 
  \right].
 \end{split}
\end{equation}
Here, $m_D \sim g T$ denotes the Debye mass (more precisely, $m_D^2=(1/3)g^2T^2(N_c+N_F/2)$ with $N_F$ light flavors). 
In the region $gT \ll |\bk| \ll T$, and also for the on-shell constraint $k^0=\bm v\cdot \bm k\ll |\bm k|$ where $\bm v$ is the heavy-quark velocity of order $\sqrt{T/M}\ll 1$, 
the spectral functions can be further simplified as \cite{LeBellac2000}
\begin{equation}
 \rho_L (k) \simeq \frac{\pi m_D^2 k^0}{|\bk|^5},
 \quad
 \rho_T (k) \simeq 
 \frac{\pi m_D^2k^0}{2 |\bk|^5},
 \label{eq:spectral-gauge}
\end{equation}
which we will use in the following calculation.

Based on these ingredients, we now evaluate the spin relaxation rate by our method.
Working out the Feynman rule for the source operator involving color magnetic field, we arrive at the following expression for the spin-traced source-source correlation function,
\begin{align}
 \chi_s\Gamma_s 
  &= {1\over 6T} \delta^{\ha\hb} G_{12}^{\Theta_{\ha} \Theta_{\hb}} (k^0 \to 0)=
  \frac{g^2}{12M^2T} 
  \delta^{\ha\hb}
 \lim_{k^0 \to 0}
 \Tr
  \left[
  \parbox{5.0cm}{\vspace{0pt}\includegraphics[width=5.0cm]{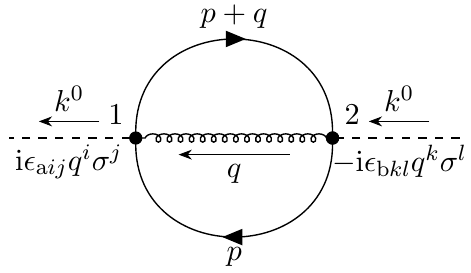}}
  \right]
  \nonumber \\
  &= \frac{(N_c^2 - 1) g^2}{6M^2 T}
  \underbrace{\int \frac{\diff^4 p}{(2\pi)^4} \frac{\diff^4 q}{(2\pi)^4} 
  S_{12} (p) S_{21} (p+q) |\bq|^2 n_B (q^0) \rho_T (q)}_{F (k^0 \to 0)}.
 \label{eq:gamma-integral}
\end{align}
Using the free propagator for heavy quarks and the HTL spectral density for the gluons, we can perform the integral as 
\begin{equation}
 \begin{split}
  F (k^0 \to 0)
  &= \int \frac{\diff^4 p}{(2\pi)^4} 
  \frac{\diff^4 q}{(2\pi)^4} 
  \big[ 1 - n_F (p^0+q^0) \big] 2 \pi \delta (p^0 + q^0 - E_{\bp + \bq})
  \\
  &\hspace{80pt} 
  \times n_F (p^0) 2\pi \delta (p^0- E_{\bp}) |\bq|^2 n_B (q^0)
  \frac{\pi m_D^2 q^0}{2 |\bq|^5}
  \\
  &= \int \frac{\diff^3 p}{(2\pi)^3} n_F (E_{\bp})
  \int \diff q^0 \frac{2\pi}{(2\pi)^3} 
  \int_{gT}^T
  \diff |\bq| |\bq|^2 \int_{-1}^1 \diff z 
  \big[ 1 - n_F (q^0 + E_{\bp}) \big]
  \\
  &\hspace{80pt} 
  \times 
  \delta \left( q^0 - \frac{\bq^2}{2M} - \frac{|\bp| |\bq|}{M} z \right)
  n_B (q^0)
  \frac{\pi m_D^2 q^0}{2 |\bq|^3},
 \end{split}
\end{equation}
where we restricted the $|\bq|$-integral in the soft regime discussed above.
Note that the $q^0$-integral is non-vanishing only in the restricted window,
\begin{equation}
 \omega_- (\bp,\bq) \leq q^0 \leq  \omega_+ (\bp,\bq) 
  \with 
  \omega_{\pm} (p,q) \equiv 
  \frac{\bq^2}{2M} \pm \frac{|\bp| |\bq| }{M}.
\end{equation}
Recalling that the fermion mass $M$ is much larger than any other scales in our problem, we find that the leading-order contribution comes from the $q^0 \to 0$ part of the integrand. 
Then, using $n_B (q^0) \simeq 1/(\beta q^0)$ and 
the dilute heavy-quark limit $1 - n_F (E_{\bp}) \simeq 1$, and performing both $z$ and $q^0$ integrals to obtain
\begin{align}
  F (k^0 \to 0)
  &\simeq 
  \frac{1}{4\pi^2}
  \int \frac{\diff^3 p}{(2\pi)^3} n_F (E_{\bp})
  \int_{gT}^T
  \diff |\bq| |\bq|^2 
  \frac{M}{|\bp| |\bq|} 
  \big[ \omega_+ (\bp,\bq) -  \omega_- (\bp,\bq) \big]
  \frac{1}{\beta} \frac{\pi m_D^2}{2 |\bq|^3}
 \nonumber \\
  &=
  \frac{T m_D^2}{4\pi}
  \int \frac{\diff^3 p}{(2\pi)^3} n_F (E_{\bp})
  \int_{gT}^T \diff |\bq|
  \frac{1}{|\bq|}
 \nonumber \\
 &= \frac{T m_D^2}{4\pi} n(T) \log (1/g),
\end{align}
where we introduced the heavy-quark number density as 
$\ds{n(T) \equiv \int \frac{\diff^3 p}{(2\pi)^3} n_F (E_{\bp})}$.
Substituting this result into eq.~\eqref{eq:gamma-integral}, 
we obtain the leading-log result as 
\begin{equation}
  \chi_s\Gamma_s 
   \simeq n(T) 
   \frac{(N_c^2 - 1) g^2 m_D^2}{24 \pi M^2} \log (1/g).
\end{equation}

The spin susceptibility $\chi_s$ can be computed by introducing a spin chemical potential 
$\mu^{\ha}$ that couples to the spin in the Hamiltonian as
$- \bm{\mu} \cdot \bm{S} (= - \mu^{\ha} S_{\ha})$ with a spin of heavy quarks $\bm{S}$.
This induces the modification of distribution functions according to spin direction as $n_F(E_{\bm q}-\bm\mu\cdot\bm S)\approx n_F(E_{\bm q})-n_F'(E_{\bm q})(\bm\mu\cdot\bm S)$.
Then, the induced net spin density along the direction of $\bm \mu$ to linear order is
\be
 - N_c \int {\diff^3 p\over (2\pi)^3}
 \sum_{S=\pm 1/2} {S} n_F' (E_{\bm p}) {S} \mu 
 = - {N_c\over 2} \int {\diff^3 p\over (2\pi)^3}
 n_F'(E_{\bm p}) \mu \equiv \chi_s \mu,
\ee
where $N_c$ is the dimension of the fundamental representation of the color gauge group $\SU(N_c)$.
As a result, we obtain the spin susceptibility $\chi_s$ as 
\be
 \chi_s = - {N_c\over 2} \int {\diff^3 p\over (2\pi)^3} n_F' (E_{\bm p}) 
 \approx 
 {N_c\over 2} \beta \int {\diff^3 p \over (2\pi)^3} n_F (E_{\bm p}) 
 ={N_c\over 2T} n(T),
\ee
where we used $n_F'(E)\approx -\beta n_F(E)$ in dilute approximation.

With this, we finally find the leading-log result of the spin relaxation rate $\Gamma_s$ as
\begin{equation}
  \Gamma_s 
   = C_2 (F) \frac{g^2 m_D^2 T}{6 \pi M^2} \log (1/g),
   \label{eq:Gamma-spin}
\end{equation}
where we introduced the Casimir operator for the heavy-quark fundamental representation $F$ as
$C_2 (F) =(N_c^2 - 1)/(2N_c)$. 
By working out color factors with a general representation, it can be shown that the result for a general representation $R$ is given by replacing $C_2(F)$ with $C_2(R)$, leading to the result presented in eq.~\eqref{eqint3}.

\subsection{Method 2: Spin density correlation function in $\omega\ll\Gamma_s$ regime}
\label{sec:spin-spin}

As explained in section~\ref{description}, a diagrammatic evaluation of the spin density correlation function in the strict hydrodynamic regime $\omega\ll\Gamma_s$ requires a summation of an infinite number of ladder diagrams that are enhanced by infrared pinching singularities. 
The summation is achieved by solving a Bethe-Salpeter (or Schwinger-Dyson) equation for the vertex corrections given by rungs in the ladder diagrams. 
In each diagram, the propagators should also include the leading imaginary part of the self-energy, which, via the optical theorem, is proportional to the total cross section of the heavy quark interacting with the background plasma, i.e., the relaxation rate. 
Thus, the computation proceeds in the same way as the procedure for the perturbative evaluation of transport coefficients~\cite{Jeon:1994if,ValleBasagoiti:2002ir,Aarts:2002tn,Hidaka:2010gh,Jimenez-Alba:2015bia}, except for the additional heavy-quark expansion. 
In the following, we divide our computation into three parts: we first derive the resummed formula for the spin density correlation, next find the Bethe-Salpeter equation for the effective vertex, and finally solve it in the heavy-quark limit.

\paragraph{1) Resummed formula for spin density correlation:} 
We choose to work in the real-time formalism in Schwinger-Keldysh contours~\cite{Schwinger:1960qe,Keldysh:1964ud}, especially in $r/a$ basis defined by $O_r={1\over 2}(O_1+O_2)$ and $O_a=O_1-O_2$, in terms of the doubled fields $O_1$ and $O_2$ living on forward and backward contours in time, 1 and 2, respectively. The retarded two point correlation function is related to $ra$ correlation function by $G_R= \rmi G_{ra}$.%
\footnote{Our definition of the retarded correlation function differs from the conventional definition in literature by a sign.}
We would like to compute $ra$ correlation function of spin density, 
$G^{J^0_{\ha} J^0_{\hb}}_{ra} \equiv \langle J^{0}_{~\ha, r} J^{0}_{~\hb,a} \rangle$, in the strict hydrodynamic regime of $\omega\to 0$, which is expected to behave as follows [recall eq.~\eqref{eq:true-hydro-spin-spin}]: 
\be
 G^{J^0_{\ha} J^0_{\hb}}_{ra}(\omega) 
 \approx 
 \left( - \rmi \chi_s + {\chi_s\over\Gamma_s}\omega + \cdots \right)
 \delta_{\ha\hb}.
 \label{Gabrs}
\ee
The pinchingd singularities affect only the real part of this
$ra$ correlation function, that is, the imaginary part of $G_R^{J^0_{\ha} J^0_{\hb}}$, which contains the spin relaxation rate $\Gamma_s$. 
We also note that the imaginary part of $G_R^{J^0_{\ha} J^0_{\hb}}$ corresponds to the spectral function of the spin density thanks to the fluctuation-dissipation relation.

Due to the presence of the pinching singularities, 
we now need to use the following resummed heavy-quark propagators in $ra$ basis in frequency-momentum space:
\begin{equation}
 \begin{split}
  S_{ra} (k)
  &= 
  \parbox{2.5cm}{\vspace{10pt}\includegraphics[width=2.5cm]{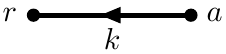}}
  = \frac{\rmi}{k^0 - E_{\bk} + \rmi \zeta_{\bk}/2},
  \\
  S_{ar} (k)
  &= 
  \parbox{2.5cm}{\vspace{10pt}\includegraphics[width=2.5cm]{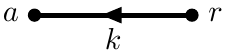}}
  = \frac{\rmi}{k^0 - E_{\bk} - \rmi \zeta_{\bk}/2},
  \\ S_{rr} (k)
  &= 
  \parbox{2.5cm}{\vspace{10pt}\includegraphics[width=2.5cm]{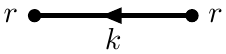}}
  = \left( \frac{1}{2} - n_F (k^0) \right)
  \left[  
   S_{ra}(k) - S_{ar}(k)
  \right],
 \end{split}
 \label{eq:resummed-fermion}
\end{equation}
where $\zeta_{\bk} \sim g^2 \log (1/g)$ denotes the leading relaxation rate for heavy quark of momentum $\bm k$, and $E_{\bm k}={\bm k^2}/(2M)$ is the non-relativistic energy. Note that the propagators are in fact  $2\times 2$ matrices in spin space, and since they are proportional to the identity matrix in non-relativistic limit, which we omit for simplicity. Once again, the Fermi-Dirac distribution at zero chemical potential in non-relativistic limit is $n_F(k^0)={1\over \rme^{\beta k^0} \rme^{\beta M}+1} \approx \rme^{-\beta M} \rme^{-\beta k^0}\ll 1$.

Recalling $J^{0}_{\ha}={1\over 2} \psi^\dagger \sigma_{\ha} \psi$, it is straightforward to draw and express real-time Feynman diagrams in the $ra$ basis~\cite{Jimenez-Alba:2015bia}. 
The summation over all ladder diagrams is achieved by solving the Bethe-Salpeter equation for the effective vertex function for $J^0_{\ha}$, which we denote as ${1\over 2}\Sigma_{\ha} (p;k)$, where $k$ is the overall momentum carried by $J^0_{\ha}$, and $p$ is the momentum of the out-going propagator from the vertex (see the figure below). 
Once the effective vertex is found by solving the Bethe-Salpeter equation, the correlation function $G^{J^0_{\ha} J^0_{\hb}}_{ra}$ is given by the following two real-time diagrams
\begin{equation}
 \begin{split}
  G^{J^0_{\ha} J^0_{\hb}}_{ra}(k)
  &= \average{J^{0}_{\ha,r} (k) J^{0}_{\hb,a} (-k)}
  \\
  &= 
  \parbox{5.0cm}{\vspace{2pt}\includegraphics[width=5.0cm]{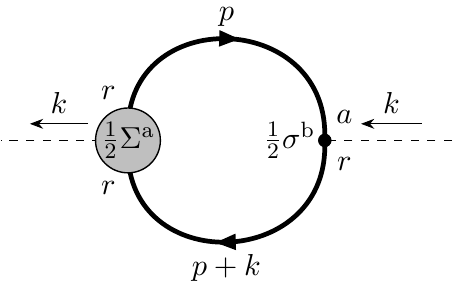}}
  ~+~ 
  \parbox{5.0cm}{\vspace{2pt}\includegraphics[width=5.0cm]{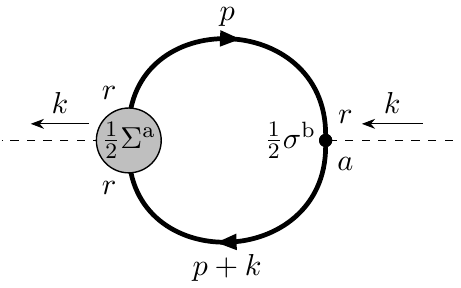}}
  \\
  &= (-1){d_R\over 4} \int \frac{\diff^4 p}{(2\pi)^4}
  \tr \left[
  \Sigma^{\ha} (p;k) S_{rr}(p+k) \sigma^{\hb} S_{ar} (p)
  + \Sigma^{\ha} (p;k) S_{ra} (p+k) \sigma^{\hb} S_{rr} (p)
  \right],
 \end{split}
\end{equation}
where $d_R$ is the dimension of the color representation of the heavy quark ($N_c$ for fundamental representation).
We recall that the pinching pole enhanced contributions arise from the combination of $S_{ra}$ and $S_{ar}$ in the parallel rail of the diagram. Therefore, we select only a part of $S_{rr}$ written as a linear superposition of $S_{ra}$ and $S_{ar}$ in eq.~(\ref{eq:resummed-fermion}), that produces a pinching singularity,
\begin{align}
  G^{J^0_{\ha} J^0_{\hb}}_{ra}(k)
  &\simeq 
 {d_R\over 4} \int \frac{\diff^4 p}{(2\pi)^4}
  \left[ n_F (p^0 + k^0) - n_F (p^0) \right]
  S_{ra} (p+k) S_{ar} (p)
  \tr 
 \big[ \Sigma^{\ha} (p;k) \sigma^{\hb} \big]
 \nonumber \\
 &\simeq
 {k^0}{d_R\over 4} 
 \int \frac{\diff^4 p}{(2\pi)^4} n_F^{\prime} (p^0)
 \frac{\rmi}{p^0 - E_{\bp} + \rmi \zeta_{\bp}/2} 
 \frac{\rmi}{p^0 - E_{\bp} - \rmi \zeta_{\bq}/2}
 \tr \big[ \Sigma^{\ha} (p;k = 0) \sigma^{\hb} \big]
 \nonumber 
 \\
 &\simeq
 -{ k^0}{d_R\over 4} 
 \int \frac{\diff^3 p}{(2\pi)^3} 
 n_F^{\prime} (E_{\bp})
 \tr \left[ 
 \frac{\Sigma^{\ha} (\bp)}{\zeta_{\bp}} \sigma^{\hb} 
 \right],
 \label{eq:spin-spin-Pinch}
\end{align}
where we take the small $k^0 \to 0$ expansion together with $\bm k\to \bm 0$, and
define $n_F^{\prime} (p^0) \equiv \diff n_F (p^0)/\diff p^0$.
To proceed the last line, we use the pinching pole approximation valid at leading order,
\be
 \frac{1}{p^0 - E_{\bp} + \rmi \zeta_{\bp}/2} 
 \frac{1}{p^0 - E_{\bp} - \rmi \zeta_{\bp}/2} 
 \longrightarrow 
 {2\pi\over \zeta_{\bp}} \delta(p^0-E_{\bp}),
 \label{eq:pinching-approx}
\ee
and also define the on-shell vertex function $\Sigma^{\ha} (\bp) \equiv \Sigma^{\ha} (p^0=E_{\bp},\bp;\bk = \bzero)$.

The rotational invariance dictates that the on-shell vertex function divided by the heavy-quark relaxation rate should take the form
\be
 \phi^{\ha} (\bp) \equiv 
 {\Sigma^{\ha} (\bp) \over\zeta_{\bp}}
 = \sigma^{\ha} f_1 (|\bp|)
 + p^{\ha} (\bp \cdot \bm \sigma) f_2 (|\bp|)
 + p^{\ha} f_3 (|\bp|) {\bm 1}_{2\times 2}
 + \epsilon^{\ha\hb\hc} p^{\hb} \sigma^{\hc} f_4(|\bp|),
\ee  
in terms of four possible scalar functions $f_i(|\bp|)~(i=1,2,3,4)$ that depend on $|\bp|$ only. Upon momentum integration in eq.~(\ref{eq:spin-spin-Pinch}), the last two terms do not contribute to $G^{J^0_{\ha} J^0_{\hb}}_{ra}$ due to angular integration, but there is another reason to expect that $f_{3,4}$ in fact vanish.
Under parity transformation, the spin vertex $\Sigma^{\ha}$ remains the same, while the momentum $\bp$ flips its sign. 
Therefore, $f_3$ and $f_4$ should be absent in a parity-even plasma without, e.g., axial chemical potential.

Inserting the above expression of $\Sigma^{\ha} (\bp)/\zeta_{\bp}$ in eq.~(\ref{eq:spin-spin-Pinch}), and replacing $p^{\ha} p^{\hb}$ with ${1\over 3} \bp^2 \delta^{\ha\hb}$ after angular integration, we arrive at the expression
\be
 G^{J^0_{\ha} J^0_{\hb}}_{ra} (\omega) = - {\omega}{d_R\over 2} 
 \left[
 \int {\diff^3 p\over (2\pi)^3} n'_F(E_{\bp}) F(|\bp|) 
 \right] \delta_{\ha\hb}, 
 \label{Gra}
\ee
where $F(|\bp|)\equiv f_1(|\bp|)+{\bp^2\over 3}f_2(|\bp|)$. Our remaining task is to find the function $F(|\bp|)$ by solving the Bethe-Salpeter equation for the summation of the ladder diagrams.

\paragraph{2) Deriving the Bethe-Salpeter equation: } 
The Bethe-Salpeter equation for $\Sigma^{\ha} (\bp)$ is an integral equation, $\Sigma^{\ha} (\bp) = \sigma^{\ha} + \Delta \Sigma^{\ha} (\bp)$, with the source $\sigma^{\ha}$ and the integral kernel $\Delta\Sigma^{\ha} (\bp)$. 
The equation is drawn diagrammatically as
\begin{equation}
  \parbox{3.0cm}{\vspace{0pt}\includegraphics[width=3.0cm]{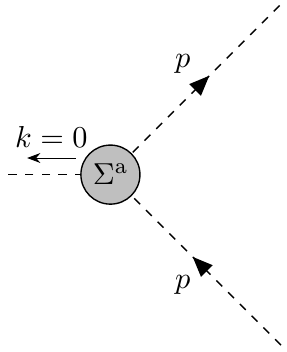}}
   ~=~
  \parbox{3.0cm}{\vspace{0pt}\includegraphics[width=3.0cm]{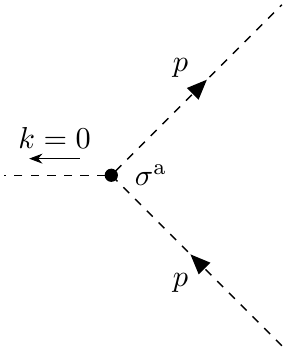}}
  ~+~
  \underbrace{
  \parbox{3.0cm}{\vspace{0pt}\includegraphics[width=3.0cm]{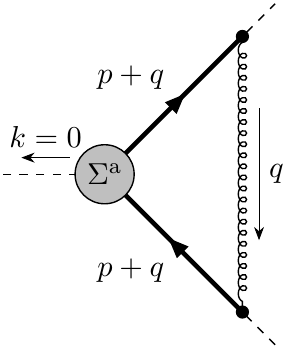}}
  }_{ \Delta \Sigma^{\ha} (\bm p)}
  \label{eq:Bethe-Salpeter}
\end{equation}
Again, focusing only on the contributions with pinching
singularities, 
we identify the following three diagrams contributing to the integral kernel $\Delta\Sigma^{\ha} (\bp)$:
\begin{equation}
 \begin{split}
  \Delta \Sigma^{\ha} (\bp)
  &=
  \parbox{3.0cm}{\vspace{0pt}\includegraphics[width=3.0cm]{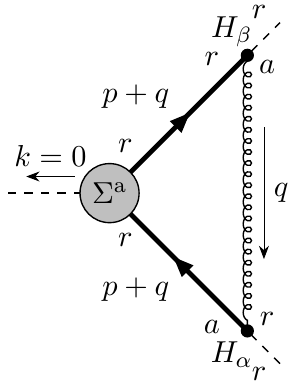}}
  + 
  \parbox{3.0cm}{\vspace{0pt}\includegraphics[width=3.0cm]{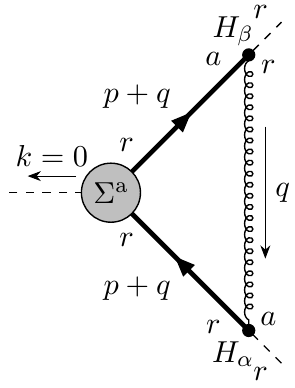}}
  ~+~
  \parbox{3.0cm}{\vspace{0pt}\includegraphics[width=3.0cm]{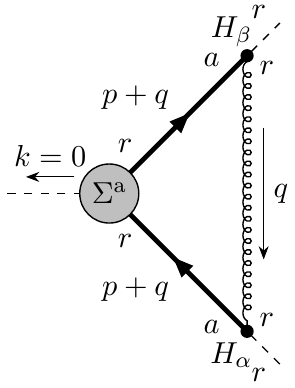}} \;\;,
 \end{split}
 \label{eq:vertex-correction}
\end{equation}
where we denote the interaction vertex of heavy quark with gluon field $A^\alpha$ as $H_\alpha(\bq)$.
Explicitly, these diagrams give us the expression for the integral kernel as 
\begin{equation}
 \begin{split}
  \Delta \Sigma^{\ha} (\bp)
  &= (-\rmi g)^2C_2(R)
  \int \frac{\diff^4 q}{(2\pi)^4}
  \Big[ 
  H_\beta (-\bq) S_{rr} (p+q) \Sigma^{\ha} (p+q) S_{ra} (p+q) 
 H_\alpha(\bq) D^{\alpha\beta}_{ra} (q)
  \\
  &\hspace{110pt}
  + H_\beta(-\bq) S_{ar} (p+q) \Sigma^{\ha} (p+q) S_{rr} (p+q) 
  H_\alpha (\bq) D^{\alpha\beta}_{ar} (q)
  \\
  &\hspace{110pt}
  + H_\beta (-\bq) S_{ar} (p+q) \Sigma^{\ha} (p+q) S_{ra} (p+q) 
  H_\alpha (\bq) D^{\alpha\beta}_{rr} (q)
  \Big],
\end{split}
\ee
where $D^{\alpha\beta} (q)=\langle A^\alpha(q)A^\beta(-q)\rangle$ is the gluon propagator in thermal equilibrium, and $C_2(R)$ is the second-order Casimir of the heavy-quark representation. 
In this expression, the interaction vertex $H_\alpha$ and the gluon propagator $D^{\alpha\beta}$ do not include color generators.

Note that the interaction vertex $H_\alpha (\bq)$ consists of three parts, 
two of which arise from the first two terms in the Lagrangian (\ref{eq:action}), respecting the $\SU(2)$ spin symmetry. They are proportional to the identity matrix $\bm{1}_{2\times 2}$.
The first term in eq.~(\ref{eq:action}) gives Coulomb interaction with $A^0$ field, and the second term gives the magnetic interaction proportional to the current, which is further suppressed by the small velocity of heavy quarks, $|\bp|/M=v\sim \sqrt{T/M}\ll 1$. 
We thus only need the leading spin conserving interaction in our computation and neglect the second term in the following.
On the other hand, we need to keep the third term in eq.~(\ref{eq:action}), which is responsible for spin relaxation in leading order. 
The interaction vertex is therefore given by
\begin{equation}
 H_\alpha (\bq) 
  = \delta_{\alpha 0} \bm{1}_{2 \times 2}
  - \frac{\rmi}{2M} \epsilon_{\alpha \hb\hc} \sigma^{\hb} q^{\hc}.
  \label{eq:heavy-quark-gluon-vertex}
\end{equation}
where the temporal component of the Levi-Civita symbol is zero: $\epsilon_{0 \hb \hc} = 0$.

We again collect only the pinching pole contributions that arise from pairing $S_{ra}$ and $S_{ar}$ for $\Delta \Sigma^{\ha} (\bp)$. 
Relying on this pinching pole approximation
and recalling
$D_{rr}^{\alpha\beta} (q)
= \left( \frac{1}{2} + n_B (q^0) \right) 
(D_{ra}^{\alpha\beta} (q) - D_{ar}^{\alpha\beta} (q))$, 
we arrive at
\begin{align}
 \Delta \Sigma^{\ha} (\bp)
 &\simeq (-\rmi g)^2 C_2(R)
 \int \frac{\diff^4 q}{(2\pi)^4}
 \big[ n_F (p^0 + q^0) + n_B (q^0) \big]
 \left[H_\beta (-\bq) \Sigma^{\ha} (p+q)  H_\alpha (\bq) \right]
 \nonumber \\
 &\hspace{104pt} \times 
 \big[ D^{\alpha\beta}_{ra} (q) - D^{\alpha\beta}_{ar} (q) \big]
 S_{ar} (p+q) S_{ra} (p+q) 
 \nonumber \\
 &\simeq g^2 C_2(R)
 \int \frac{\diff^4 q}{(2\pi)^4}  
 \big[ n_F (q^0+E_{\bp}) + n_B (q^0) \big]
 \left[ 
 H_\beta (-\bq) \frac{\Sigma^{\ha} (\bp+\bq)}{\zeta_{\bp+\bq}} H_\alpha (\bq)
 \right] 
 \rho^{\alpha\beta} (q)
 \nonumber \\  
 &\hspace{86pt} 
 \times 2\pi \delta( q^0 + E_{\bp} - E_{\bp + \bq}),
\end{align}
where $D^{\alpha\beta}_{ra} (q) - D^{\alpha\beta}_{ar} (q) = \rho^{\alpha\beta} (q)$ is the gluon spectral density, for which we use the HTL result summarized in eqs.~\eqref{eq:spectral-gluon-decomposition}-\eqref{eq:spectral-gauge} for our leading-log computation. 
We also use the on-shell condition $p^0=E_{\bp}$, that is imposed in the evaluation of $G^{J^0_{\ha} J^0_{\hb}}_{ra}$ in eq.~(\ref{eq:spin-spin-Pinch}). 

In terms of the function $\phi^{\ha} (\bp) \equiv \Sigma^{\ha}(\bp)/\zeta_{\bp}$, which can be shown to correspond to a linearized distribution function in kinetic theory, the integral equation depicted in eq.~\eqref{eq:Bethe-Salpeter} then takes the following form
\begin{align}
 - \sigma^{\ha} 
 &= - \zeta_{\bp} \phi^{\ha}(\bp) 
 + g^2 C_2(R) 
 \int \frac{\diff^4 q}{(2\pi)^4}  
 \big[ n_F (q^0+E_{\bp}) + n_B (q^0) \big]
 \left[H_\beta (-\bq) \phi^{\ha} (\bp + \bq) H_\alpha (\bq)\right] 
 \rho^{\alpha\beta} (q)
 \nonumber \\ 
 &\hspace{142pt} 
 \times 2\pi \delta(q^0 + E_{\bp}-E_{\bp + \bq}).
 \label{eq:integral-eq}
\end{align}
The important observation, that is common in the similar types of integral equations appearing in the computation of transport coefficients, is that the relaxation rate $\zeta_{\bp}$ from the imaginary part of one-loop self-energy diagram is computed by a very similar expression to the integral kernel, 
\begin{equation}
 \begin{split}
  \zeta_{\bp} 
  &= {1\over 2} g^2 C_2(R)
  \int \frac{\diff^4 q}{(2\pi)^4}   
  \big[ n_F (q^0 + E_{\bp}) + n_B (q^0) \big]
  \tr
  \left[ H_\beta (-\bq) H_\alpha (\bq) \right] 
  \rho^{\alpha\beta} (q)
  \\ 
  &\hspace{90pt}
  \times 2\pi \delta(q^0 + E_{\bp} - E_{\bp + \bq}),
 \end{split}
\end{equation}
where the factor $1/2$ in front comes from the average over spin states since the relaxation rate is independent of the spin states due to rotational invariance. 
This allows us to combine the two terms in the right-hand side of the integral equation~\eqref{eq:integral-eq}, and we finally arrive at the Bethe-Salpeter equation for $\phi^{\ha}$ as
\begin{align}
 -\sigma^{\ha}
 &= g^2 C_2(R)
 \int \frac{\diff^4 q}{(2\pi)^4} 
 \left(
 H_\beta (-\bq) \phi^{\ha} (\bp + \bq) H_\alpha (\bq)
 - {1\over 2} 
 \tr \left[H_\beta (-\bq) H_\alpha (\bq) \right]
 \phi^{\ha} (\bp) 
 \right) 
 \rho^{\alpha\beta} (q)
 \nonumber \\
 &\hspace{90pt}
 \times 
 \big[ n_F (q^0 + E_{\bp}) + n_B (q^0) \big]  
 2\pi \delta(q^0 + E_{\bp}-E_{\bp + \bq}).
 \label{eq:fininteq}
\end{align}
In the following, we will evaluate this integral equation in leading-log approximation by expanding in powers of small $\bq \sim gT$ to quadratic order to derive a second-order differential equation for $\phi^{\ha} (\bp)$ in $\bp$ space. 
This can be considered as the linearized collision term in quantum kinetic theory, perturbed by an external source for the spin density given by the left-hand side of the integral equation.

Before moving to the evaluation of $\Gamma_s$, it is instructive to consider a limit where the $\SU(2)$ spin symmetry is respected, by assuming that $H_\alpha(\bq)$ contains only the leading term in $1/M$, i.e., the identity matrix in spin space. In this case, the integral equation becomes
\begin{align}
-\sigma^{\ha} 
 &\simeq 
 g^2 C_2(R) \int \frac{\diff^4 q}{(2\pi)^4} 
 \big[ 
 \phi^{\ha} (\bp + \bq) - \phi^{\ha} (\bp) \big]
 \rho_L (q)
 \nonumber \\
 &\hspace{84pt}
 \times
 \big[ n_F (q^0+E_{\bp}) + n_B (q^0) \big] 
 2\pi \delta( q^0 + E_{\bp} - E_{\bp + \bq} ).
\end{align}
To see what happens in this limit, we integrate both sides in $\bp$-space after multiplying a factor 
$W(\bp) = n_F(E_{\bp}) [1 - n_F (E_{\bp})]$. 
By suitable shift of variables and using the properties of the spectral density 
$\rho (q^0,\bq)= - \rho(-q^0,\bq)=\rho(q^0,-\bq)$, 
one can show that the right-hand side vanishes after using the identity
\be
 W (\bp + \bq) 
 \big[ 
 n_F (E_{\bp}) + n_B(E_{\bp} - E_{\bp + \bq}) 
 \big]
 + W (\bp) 
 \big[ 
 n_F(E_{\bp + \bq}) + n_B (E_{\bp +\bq} -E_{\bp} )
 \big]
 = 0.
\ee
This parallels the conservation property of total charges in the collision term in kinetic theory.
Since $W(\bp)>0$, the left-hand side is not zero after $\bp$-integration, which means that the only solution for $\phi^{\ha} (\bp)$ in this limit is $\phi^{\ha} \to \infty$. 
This would imply $G_{ra}^{J^0_{\ha} J^0_{\hb}}\sim {\chi_s\over\Gamma_s}\omega \delta_{\ha\hb}\to\infty$, which correctly gives us $\Gamma_s \to 0$ in the limit.
From this consideration, it becomes clear that a finite value of $\Gamma_s$ results only from the spin violating interaction in the vertex $H_\alpha(\bq)$ due to the Pauli term.
This should be closely related to the (non-)conservative Ward-Takahashi identity for spin density. 

\paragraph{3) Solving the Bethe-Salpeter equation in heavy-quark limit: }
The most laborious part of our computation in this method is the leading-log evaluation of the integral equation (\ref{eq:fininteq}).
We first note the presence of a scale hierarchy in the current problem.
The leading-log contributions come from the soft momentum exchange $gT \ll |\bq| \ll T$, and the on-shell condition implies $q^0=E_{\bp +\bq} - E_{\bp} \approx \bv_{\bp} \cdot \bq \ll |\bq|$
with $\bv_{\bp} \equiv \bp/M$
, where we use the fact that the typical momentum of heavy quark in eq.~(\ref{eq:spin-spin-Pinch}) is $\bp \sim \sqrt{MT}$ from $E_{\bp}\sim T$. 
This gives us a hierarchy of scales $q^0 \ll |\bq| \ll |\bp|$. 
In the integral equation, we, therefore, expand each term in powers of $|{\bq}|/|\bp|\sim g$, and higher-order terms in $\bq$ and $q^0$ bring about terms of higher powers of coupling constant $g$. 
To our leading-log computation, it is thus sufficient to expand the terms in the integral equation to second order in $\bq$ and $q^0$~\cite{ValleBasagoiti:2002ir}. 

For example, recalling the spin structure of $\phi^{\ha}(\bp)$ consistent with rotational symmetry, $\phi^{\ha}(\bp) = \sigma^{\ha} f_1 (|\bp|) + p^{\ha} (\bp \cdot \bm \sigma) f_2 (|\bp|)$, the expansion of $f_1(|\bp + \bq|)$ would be
\be
 f_1(|\bp + \bq|) 
 \approx 
 f_1(|\bq|)
 + \left( 
    |\bq| \cos \theta 
    +{|\bq|^2\over 2 |\bp|} \sin^2 \theta 
   \right) 
 f'_1(|\bp|)
 + {1\over 2}|\bq|^2 \cos^2 \theta f''_1(|\bp|) ,
\ee
where $\theta$ is the angle between $\bp$ and $\bq$. Similarly, we need for our purpose
\be
 n_F (q^0 + E_{\bp}) + n_B (q^0)
 \approx 
 n_F(E_{\bp}) + {1\over\beta q^0} - {1\over 2} + {\cal O}(q^0).
\ee
Furthermore, we can safely neglect $n_F(E_{\bp})$ in the following since it is given by $n_F(E_{\bp}) \approx \rme^{-\beta M} \rme^{-\beta E_{\bp}}\ll 1$ in the dilute limit of heavy quarks.
It is also useful to perform the integration over the angle $\theta$ first, by making the energy $\delta$-function into the one for the angle $\theta$,
\be
 \delta (q^0 + E_{\bp} - E_{\bp + \bq})
 = {M\over |\bp||\bq|} \delta 
 \left(
 \cos \theta - {M\over |\bp||\bq|}
 \left( q^0 - {|\bq|^2\over 2M} \right)
 \right),
\ee
together with restricting the integration range for $q^0$ as
\be
 {|\bq|^2\over 2M} - {|\bp||\bq|\over M} 
 \leq q^0 \leq 
 {|\bq|^2\over 2M} +{|\bp||\bq|\over M}.
 \ee
Finally, we use the gluon spectral density in the Coulomb gauge
\be
 \rho^{\alpha\beta} (q) 
 = \delta^{\alpha 0} \delta^{\beta 0} \rho_L(q) 
 + P_T^{\alpha\beta}(\bq) \rho_T (q),
\ee
where $P_T^{ij}(\bq) = \delta^{ij}- q^i q^j/|\bq|^2$ is the projection operator for spatial components $i$ and $j$. 
The HTL resummed longitudinal and transverse spectral densities in the regime $q^0 \ll |\bq| \ll T$ become the ones given in eq.~(\ref{eq:spectral-gauge}), which is sufficient for our leading-log computation.

With all these ingredients used in the integral equation \eqref{eq:fininteq}, performing the leading-log integral for $q$ and comparing the spin structures in both sides of the equation, we arrive at the two coupled second-order differential equations for $f_1$ and $f_2$ as 
\begin{subequations}
\begin{align}
  {1\over 2} f''_1(|\bp|)
  + {1\over |\bp|}{f'_1(|\bp|)}
  + f_2(|\bp|)
  - {|\bp|\over 2TM} f'_1(|\bp|)
  -{1\over M^2}f_1(|\bp|)
  &=-{1\over T^2 \Gamma },
  \\
  {1\over 2 }f_2''(|\bp|)
  + {3\over |\bp|} f'_2(|\bp|)
  - {|\bp|\over 2TM}\left(f'_2(|\bp|)
  + {2\over |\bp|}f_2(|\bp|)\right)
  - {1\over M^2}f_2(|\bp|)
  &=0,
\end{align}
\end{subequations}
where we define 
\be
 \Gamma\equiv{C_2(R)}{g^2 m_D^2\log(1/g)\over 6\pi T}\sim g^4\log(1/g)T.
\ee
Interestingly, the two equations can be combined to give a single differential equation for $F(|\bp|)=f_1(|\bp|)+{1\over 3}|\bp|^2 f_2(|\bp|)$, that is precisely what we need in order to evaluate the correlation function $G^{J^0_{\ha} J^0_{\hb}}_{ra}$ in eq.~(\ref{Gra}),
\be
 {1\over 2}F''(|\bp|)
 + {1\over |\bp|}F'(|\bp|)
 - {|\bp|\over 2TM} F'(|\bp|)
 - {1\over M^2} F(|\bp|) 
 = -{1\over T^2\Gamma},
\label{eq:diff-eq}
\ee
which can be analytically solved in the heavy-quark limit.

Before solving this equation in $1/M$ expansion, it is a useful check to see that the first three terms, that arise from the leading spin-conserving Coulomb interaction, reproduces the conservation property of the collision term in kinetic theory, 
\be
 \int \diff^3 p \,\,
 W(\bp) 
 \left(
 {1\over 2}F''(|\bp|) 
 + {1\over |\bp|}F'(|\bp|)
 - {|\bp|\over 2TM} F'(|\bp|)
 \right)
 =0,
\ee
where the weight factor in our dilute limit becomes $W(\bp) = n_F(E_{\bp}) [1-n_F(E_{\bp})] \sim \rme^{-\beta E_{\bp}} = \rme^{-\bp^2/(2TM)}$, up to a constant factor.
Therefore, without the last term, which violates the conservation of spin, the solution would diverge $F\to\infty$. 

This implies that the solution has the following expansion in $1/M$ as
\begin{equation}
 F (|\bp|) = 
 {M^2}F_{(0)} (|\bp|) + F_{(1)} (|\bp|) 
 + {\cal O}(M^{-2}),
\end{equation}
and we are now only interested in the leading solution $F^{(0)}$.
Inserting this expansion into the differential equation~\eqref{eq:diff-eq}, 
we obtain the hierarchy of equations
\begin{subequations}
\begin{align}
  {1\over 2} F_{(0)}'' (|\bp|)
  + {1\over |\bp|} F_{(0)}' (|\bp|)
  - {|\bp| \over 2TM} F_{(0)}' (|\bp|) 
  &=0,
  \\
  {1\over 2} F_{(1)}'' (|\bp|) 
  + {1\over |\bp|} F_{(1)}'(|\bp|)
  - {|\bp|\over 2TM} F_{(1)}' (|\bp|) 
  - F_{(0)}(|\bp|) 
  &= - {1\over T^2\Gamma}.
 \end{align}
\end{subequations}
It is easy to find the solution $F_{(0)}(|\bp|)=C$ by inspection, where $C$ is a constant. To determine $C$, we integrate both sides of the second equation with the weight factor $W(\bp) = \rme^{-\beta E_{\bp}}$, that removes the term with $F^{(1)}$, and we arrive at $C=1/(T^2\Gamma)$.
With this leading solution 
\be 
 F(|\bp|)={M^2\over T^2\Gamma}+{\cal O}(M^0),
\ee
together with eq.~(\ref{Gra}), we finally arrive at the correlation function 
\be
 G^{J^0_{\ha} J^0_{\hb}}_{ra}(\omega)
 = -{\omega}{M^2\over T^2\Gamma}
 \left[
  {d_R\over 2}\int {\diff^3 p\over (2\pi)^3} 
  n_F'(E_{\bp})
 \right] \delta_{\ha\hb},
 \label{finalG}
\ee
which should be identified as $\omega {\chi_s\over\Gamma_s}\delta_{\ha\hb}$, according to eq.~(\ref{Gabrs}).
Using
the spin susceptibility $\chi_s$ 
\be
 \chi_s =
 - {d_R\over 2} \int{\diff^3 p\over (2\pi)^3}
 n_F'(E_{\bp}),
 \ee
our result (\ref{finalG}) finally gives us the spin relaxation rate $\Gamma_s$
\begin{equation}
 \Gamma_s=\Gamma{T^2\over M^2}={C_2(R)}{g^2 m_D^2\log(1/g) T\over 6\pi M^2},
  \label{eq:Gamma-spin2}
\end{equation}
which precisely coincides with eq.~\eqref{eq:Gamma-spin} obtained in the previous subsection.

\subsection{Method 3: Quantum kinetic theory for spin in heavy-quark limit}
\label{sec:kinetic}

Kinetic theory is an intuitive and powerful framework for describing time-dependent dynamics of weakly interacting quasi-particles in phase space (position-momentum space)~\cite{Landau:Kinetic}. 
It is a statistical description of the system in terms of particle number distribution in phase space. 

When the two spin states are nearly degenerate in energy, and the quantum correlation time between two spin states is comparable to a macroscopic time scale of interests, one should instead consider the full $2\times 2$ density matrix distribution in spin space, $\hat\rho(\bm x,\bm p,t)$, to properly describe time-dependent dynamics of spin polarization of quasi-particles, that is, the quantum kinetic theory. 
The $2\times 2$ density matrix in spin basis can be decomposed into the form 
\be
 \hat\rho(\bm x,\bm p,t) = {1\over 2}f(\bm x,\bm p, t) \bm{1}_{2 \times 2} + \bm S(\bm x,\bm p,t)\cdot\bm\sigma,
\ee 
where $f(\bm x,\bm p,t)= \tr [\hat\rho(\bm x,\bm p,t)]$ is identified as the usual particle number distribution function, while $\bm S(\bm x,\bm p,t)= \tr \left[\hat\rho(\bm x,\bm p,t){\bm\sigma\over 2}\right]$ has the interpretation of the spin distribution function. The time evolution equation for each of them may be written in a similar form as the Boltzmann equation with proper collision terms. Alternatively, the equation written in terms of $\hat\rho$ itself should take a form of the Lindblad equation, in general.

In ref.~\cite{Li:2019qkf}, the quantum Boltzmann equation of $f(\bm p,t)$ and $\bm S(\bm p,t)$ for massive quarks has been constructed to leading-log order of QCD coupling constant, when they are spatially homogeneous, 
\be
 {\partial f(\bm p,t)\over\partial t} = 
 \hat{\Gamma}_f [ f(\bm p,t)]\,,
 \quad {\partial \bm S(\bm p,t)\over\partial t}
 =\hat{\bm\Gamma}_S[ \bm S(\bm p,t)],\label{Boltzmann}
\ee
where the collision terms, or the relaxation operators, 
$\hat{\Gamma}_f$ and $\hat{\bm \Gamma}_S$, are universally of order $g^4\log(1/g)T$. 
They take a form of second order differential operator in momentum space $\bm p$. 
The $\hat{\Gamma}_f$ is nothing but the usual collision term for quark number distribution function in QCD.
The collision term for the spin polarization, i.e. $\hat{\bm \Gamma}_S$, is a novel object, whose explicit form in spatially homogeneous limit can be found in ref.~\cite{Li:2019qkf}. 

We are interested in the heavy-quark limit, i.e., $M\gg T$, of $\hat{\bm\Gamma}_S$, that determines the relaxation dynamics of heavy-quark spin in momentum space, which ultimately gives us the spin relaxation rate $\Gamma_s$ in the spin hydrodynamic regime. 
Although it is possible to take $M\gg T$ limit of the result in ref.~\cite{Li:2019qkf}, we present in this section the derivation of $\hat{\bm \Gamma}_S$ directly from our non-relativistic effective theory (see also the derivation based on the Kadanoff-Baym formalism in appendix \ref{sec:Kadanoff-Baym}). 
For clarity of the presentation, we will keep only the leading spin-conserving Coulomb interaction and the leading spin-violating Pauli-interaction, as in the previous section. 
This is sufficient to obtain the correct spin relaxation rate at the leading order. 
For more details of our method of derivation, we refer to ref.~\cite{Li:2019qkf}.

The full density operator 
of heavy quark in spatially homogeneous limit is written in momentum basis as
\be
 \hat\rho(t)=\int {\diff^3 p\over(2\pi)^3}\hat \rho(\bm p,t),
\ee
where $\hat\rho(\bm p,t)$ is the density matrix in spin space in the subspace of a given momentum $\bm p$. 
More explicitly, we can express it as $\hat\rho(\bm p,t)=\sum_{s,s'}|\bm p,s\rangle\rho_{ss'}(\bm p,t)\langle \bm p,s'|$, where $|\bm p,s\rangle$ is the eigenstate of momentum $\bm p$ and spin basis state $s$. The $2\times 2$ matrix $\rho_{ss'}(\bm p,t)$ is equivalent to the Wigner function of non-relativistic two-component quark field operators of spin 1/2, and our method in the following 
gives the same result as in the Kadanoff-Baym approach at leading order as shown in appendix \ref{sec:Kadanoff-Baym}.
 
The density matrix evolves in time by quantum mechanical unitary evolution in the thermal QCD plasma,
\be
\hat\rho(t+\Delta t)=\bigg\langle\hat U_1(t+\Delta t,t)\hat\rho(t)\hat U_2^\dagger(t+\Delta t,t)\bigg\rangle_A,
\ee
where $\langle\cdots\rangle_A$ denotes the thermal average of background gluon fields interacting with the heavy quark, and the subscript $1$ and $2$ refers to as the forward and backward Schwinger-Keldysh contours, respectively. 
This is because the forward contour 1 describes the evolution of ket state $|\bm p,s\rangle$, while the backward contour 2 is for the conjugate bra state $\langle\bm p,s'|$. 
Note also that the gluon fields in the evolution operator $\hat U_{1,2}$ are the Schwinger-Keldysh fields $A^{1,2}_\mu$ on the contours 1 and 2, respectively, whose correlation functions satisfy the fluctuation-dissipation relations in thermal equilibrium.

Expanding $\hat U(t+\Delta t,t)$ up to the second-order of the quark-gluon interactions in the interaction picture and performing a thermal average of gluon two-point correlation functions, one obtains the time evolution equation for the density matrix as
\begin{equation}
 {\partial \hat\rho(\bm p,t)\over\partial t}
 = g^2 C_2(R) (\Gamma_{\rm cross} + \Gamma_{\rm self\,energy}),
 \label{eq:eom-density-operator}
\end{equation}
where $\Gamma_{\rm cross}$ arises from the correlation of first order terms in quark-gluon interaction in contours 1 and 2, while $\Gamma_{\rm self\,energy}$ is from the second order terms in each contour separately. 
The diagrammatic representation is depicted in Figure \ref{fig:Gamma}.
\begin{figure}[t]
 \centering
 \includegraphics[width=0.9\linewidth]{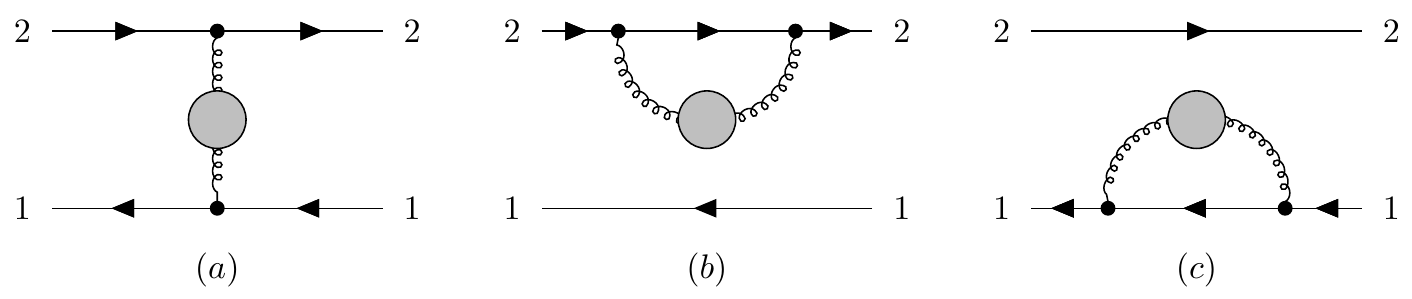}
 \caption{ The ``cross'' contribution (a), and the two self-energy contributions (b) and (c). In these figures, time flows from the right to the left.
 To emphasize the HTL contribution in the gluon propagator, we put the blobs in these figures.} 
 \label{fig:Gamma}
\end{figure}
Explicitly, we find
\begin{align}
 \Gamma_{\rm cross} &= 
 \int {\diff^4 q \over (2\pi)^4} 
 H_{\alpha}(\bm q) 
 \hat\rho(\bm p-\bm q,t) 
 H_\beta(-\bm q) 
 D_{12}^{\alpha\beta} (q)
 2\pi \delta(q^0+E_{\bm p-\bm q}-E_{\bm p}),
 \\
 \Gamma_{\rm self\,energy} 
 &= -{1\over 2} \int{\diff^4 q\over (2\pi)^4} 
 \tr \big[ H_\alpha(\bm q)H_\beta(-\bm q) \big]
 D^{\alpha\beta}_{21} (q) 
 2\pi \delta(q^0+E_{\bm p-\bm q}-E_{\bm p}) \hat\rho(\bm p,t),
 \label{eq:damping-term-density-operator}
\end{align}
where $D_{12}^{\alpha\beta} (q)$ and $D_{21}^{\alpha\beta} (q)$ are 
the thermal gluon correlation functions on Schwinger-Keldysh contours defined in eq.~\eqref{eq:prop}, and 
$H_\alpha (\bq)$ is the heavy quark-gluon interaction vertex given in eq.~\eqref{eq:heavy-quark-gluon-vertex}.
In deriving the self-energy term, we need to use the identity 
$D_{11}(t)\theta(t) + D_{22}(t)\theta(-t) = D_{21}(t)$, 
and also the fact that the self-energy is proportional to identity matrix in spin space due to parity symmetry.

One can understand the structure of the above equation~\eqref{eq:eom-density-operator} from the viewpoint of the Lindblad equation~\cite{Lindblad:1975ef}: 
\begin{equation}
 {\partial\hat\rho\over\partial t} 
  = \sum_i \gamma_i \hat\rho \gamma_i^\dagger
   - {1\over 2}
   \{\gamma_i^\dagger\gamma_i,\hat\rho\}_+.
\end{equation}
with the anti-commutator 
$\{A,B\}_+ \equiv AB + BA$.
The self-energy term corresponds to a dissipative loss of probability, while the cross term is a gain term that ensures the total probability conservation. The index $i$ becomes the $t$-channel gluon momentum $q$. 
One can explicitly check that the total trace of the density matrix in momentum-spin space is indeed conserved in our evolution equation.
It is also possible to show at this stage that the Boltzmann equilibrium density matrix, 
$\hat\rho_{\rm eq}(\bm p) \propto {\bf 1}_{2\times 2} \rme^{-\beta E_{\bm p}}$, 
is a zero mode of the collision operator.

The computation of the above $\Gamma_{\rm cross}$ and $\Gamma_{\rm self\,energy}$ in leading-log order is very similar to that in the previous section. Since we are interested in the spin polarization of the density matrix, we focus only on the spin part, $\hat\rho(\bm p)\sim \bm S(\bm p)\cdot\bm\sigma$, in the following.
Using the gluon spectral density \eqref{eq:spectral-gluon-decomposition} in Coulomb gauge, a simple algebra leads to
\begin{equation}
 {\partial\rho(\bm p,t)\over\partial t}
  = g^2 C_2(R) \int{\diff^4 q\over (2\pi)^4} 
  \left[
   C_L \,\rho_L (q) - {1\over 2M^2}C_T\,\rho_T (q)
  \right]
  2\pi \delta (q^0+E_{\bm p-\bm q}-E_{\bm p}),
\end{equation}
where we introduced
\begin{equation}
 \begin{split}
  C_L &= 
  \big[ \bm S(\bm p-\bm q)-\bm S(\bm p) \big] \cdot \bm\sigma 
  \,\,n_B(q^0)-\bm S(\bm p) \cdot \bm\sigma,
  \\
  C_T &= 
  \big[ 
  (\bm q\cdot\bm\sigma) \big(\bm q\cdot\bm S(\bm p-\bm q) \big)  
  + \bm q^2 \bm S(\bm p)\cdot\bm\sigma 
  \big] n_B(q^0)
  + \bm q^2 \bm S(\bm p) \cdot \bm\sigma.
 \end{split}
\end{equation}
Note that the $C_L$ comes from the Coulomb interaction respecting 
the spin symmetry, while $C_T$ does from the Pauli interaction that breaks it.

As in the previous sections, we perform the angular integration of $\cos\theta$ between $\bm p$ and $\bm q$ first, using the relation $\cos\theta={M\over |\bm p|}{q^0\over|\bm q|}+{|\bm q|\over 2|\bm p|}$.
This replaces the $q$ integration as 
\begin{equation}
 \int{\diff^4 q\over (2\pi)^4} 
  2\pi \delta (q^0+E_{\bm p-\bm q}-E_{\bm p}) 
  \to {1\over (2\pi)^2}{M\over |\bm p|}
  \int_{gT}^T \diff |\bm q| |\bm q| \int \diff q^0,
\end{equation}
with the $q^0$-integration range
$-{|\bm q|^2\over 2M} -{|\bm p||\bm q|\over M} \leq q^0 \leq -{|\bm q|^2\over 2M} + {|\bm p||\bm q|\over M}$.
In computing the $\bm q$-integration, we have to use $\bm q=\bm q_L+\bm q_T=|\bm q|\cos\theta\hat{\bm p}+\bm q_T$ with $\cos\theta$ given as above, while the $\bm q_T$-integration gives zero after polar angle integration. 
Similarly, $q^i q^j$ is replaced by 
\begin{equation}
 q^i q^j \to 
  |\bm q|^2 
  \left[
   \cos^2 \theta \hat{p}^i \hat{p}^j 
   + {1\over 2} \sin^2 \theta 
   \left( \delta^{ij} - \hat{p}^i \hat{p}^j \right)
  \right] .
\end{equation}
The leading-log arises from the expansion of $C_{L,T}$ up to second order in $\bm q\sim gT\ll \bm p$, and performing leading-log integral in $|\bm q|$, with the range from $m_D\sim gT$ to the upper bound of $T$, where our expression of HTL gluon spectral density breaks down.

The result for $\hat{\bm\Gamma}_S$ is organized as the $1/M$ expansion, 
\begin{equation}
 \hat{\bm\Gamma}_S = 
 \hat{\bm\Gamma}_S^{(0)} + \hat{\bm\Gamma}_S^{(1)} + \cdots,
\end{equation}
where $\hat{\bm\Gamma}_S^{(0)}$ arises from the longitudinal Coulomb interaction, $C_L$, while the next order term $\hat{\bm\Gamma}_S^{(1)}$ of order $1/M^2$ results from the Pauli-interaction, $C_T$.
The leading term $\hat{\bm\Gamma}_S^{(0)}$, which conserves the total spin density, turns out to be identical to the leading term in $\hat{\Gamma}_f$, which describes the momentum diffusion of heavy quark with the known heavy-quark drag force coefficient $\eta_D$ as~\cite{Moore:2004tg}
\begin{equation}
 \hat{\Gamma}_S^{(0)\ha}[\bm S(\bm p)]
  = \eta_D \bm\nabla_{p}\cdot\left(\bm p \,S^\ha+TM\bm\nabla_{p}  S^\ha\right)
  \with
  \eta_D = C_2(R) {g^2 m_D^2\log(1/g)\over 12\pi M}
\end{equation}
This results in the Fokker-Planck equation for the density matrix equivalent to the Langevin equation of heavy-quark Brownian motion, with each component $S^\ha(\bm p)$ playing a role of distribution in momentum space,
\begin{equation}
 \dot{\bm p}=-\eta_D \bm p+\bm\xi(t)\,,\quad \langle \xi^\ha(t)\xi^\hb(t')\rangle=\kappa \delta^{\ha\hb}\delta(t-t'),
\end{equation}
with $\kappa=2MT \eta_D$ from the fluctuation-dissipation relation. 
Physically, what happens is that the spin attached to the heavy quark simply follows the motion of the heavy quark in momentum space. The same result has also been obtained to this order in ref.~\cite{Yang:2020hri}.
The next leading collision operator, $\hat{\bm\Gamma}_S^{(1)}$, which is new and encodes the spin-violating effect, is given by a simple expression
\begin{equation}
 \hat{\bm\Gamma}_S^{(1)} [\bm S(\bm p)] 
  = - \eta_D{2T\over M}\bm S(\bm p) 
  =-C_2(R) {g^2 m_D^2T\log(1/g)\over 6\pi M^2}\bm S(\bm p),
  \label{eq:Gamma-S1}
\end{equation}
from which we can evaluate the spin relaxation rate in the leading-log.

The leading $\hat{\bm\Gamma}_S^{(0)}$ term determines the relaxation rates of non-hydrodynamic modes of spin distribution in momentum space. 
In other words, its non-zero eigenvalue corresponds to the relaxation rate of the non-hydrodynamic eigenmode. 
The only exception is the mode with zero eigenvalue, which corresponds to the quasi-hydrodynamic 
(or Hydro+) mode of spin density. 
It is easy to guess what the zero mode should be: it is the equilibrium Boltzmann distribution, $\bm S_0 \rme^{-\beta E_{\bm p}}$, with any constant vector $\bm S_0$. 
Indeed, it is a simple algebra to see that $\hat{\bm\Gamma}_S^{(0)}[\bm S_0 \rme^{-\beta E_{\bm p}}]=0$, while the next order term $\hat{\bm\Gamma}_S^{(1)}$ gives the relaxation dynamics of spin that we are interested in.

After integration over $\bm p$, the spin density in position space is given by 
\begin{equation}
 \bm S = 
  \int{\diff^3 p\over (2\pi)^3}
  \bm S_0 \rme^{-\beta E_{\bm p}}
  = \left(\frac{TM}{2\pi} \right)^{3/2} \bm S_0,\label{463}
\end{equation}
and $\bm S_0$ can be identified as the quasi-hydrodynamic mode of spin density. Therefore, we are led to write down the spin distribution function in 
the $1/M$ expansion as
\begin{equation}
 \bm S(\bm p,t) = 
  \bm S_0(t) \rme^{-\beta E_{\bm p}} + \Delta \bm S(\bm p,t),
  \label{hydroexp}
\end{equation}
where $\Delta \bm S(\bm p,t)$ contains all non-hydrodynamic modes which relax much faster than $\bm S_0(t)$. It is defined by requiring $\bm S(\bm p,t)$ and $\bm S_0(t)$ to satisfy the matching condition (\ref{463}),
which means that $\int_{\bm p} \Delta \bm S(\bm p,t)=0$. 
It is expected that $\Delta \bm S(\bm p,t)$ is smaller than the leading term by the hydrodynamic expansion parameter $\omega/\Gamma$. 
In the regime of spin hydrodynamics where $\omega\sim \Gamma_s\sim (T/ M)^2\Gamma$, 
the derivative expansion becomes equivalent to $T/M\ll 1$ expansion.
The relaxation rate of the quasi-hydrodynamic mode, $\bm S_0(t)$, 
is the spin relaxation rate $\Gamma_s$ that we are interested in.

After using the expansion (\ref{hydroexp}) in the quantum Boltzmann equation (\ref{Boltzmann}),
and integrating over $\bm p$, we arrive at
\begin{equation}
 {\partial \bm S_0(t)\over\partial t}
  = {\int_{\bm p}\hat{\bm\Gamma}_S^{(1)}[\bm S_0 \rme^{-\beta E_{\bm p}}]\over\int_{\bm p} \rme^{-\beta E_{\bm p}}},
  \label{s0t}
\end{equation}
where we used the matching condition (\ref{463}), as well as the conservation property $\int_{\bm p} \hat{\bm\Gamma}_S^{(0)}[\bm S(\bm p)]=0$ for any $\bm S(\bm p)$. 
We dropped the higher-order terms in the $1/M$ and small frequency expansion, e.g.,
by neglecting the term such as
$\hat{\bm\Gamma}_S^{(1)}[\Delta\bm S(\bm p)]$. 
With our result of $\hat{\bm\Gamma}_S^{(1)}$ given in eq.~\eqref{eq:Gamma-S1}, we have
\begin{equation}
 \hat{\bm\Gamma}_S^{(1)}[\bm S_0 \rme^{-\beta E_{\bm p}}]
  = -{\eta_D}{2 T\over M}\bm S_0 \rme^{-\beta E_{\bm p}},
\end{equation}
which enables us rewrite eq.~(\ref{s0t}) as the following simple relaxation equation:
\begin{equation}
 {\partial \bm S_0(t)\over\partial t} = -\Gamma_s \bm S_0(t).
\end{equation}
Here, we eventually identify the spin relaxation rate $\Gamma_s$ from the quantum kinetic theory as
\begin{equation}
 \Gamma_s =
  {\eta_D}{2 T\over M}{\int_{\bm p} \rme^{-\beta E_{\bm p}}\over \int_{\bm p} \rme^{-\beta E_{\bm p}}}
  = {\eta_D}{2 T\over M}
  = {C_2(R)}{g^2 m_D^2\log(1/g) T\over 6\pi M^2},
  \label{eq:Gamma-spin3}
\end{equation}
which agrees with the result of the diagrammatic methods in the previous subsections [see eqs.~\eqref{eq:Gamma-spin} and  \eqref{eq:Gamma-spin2}].

\section{Summary and Outlook} 
\label{sec:summary}

In this paper, we have evaluated the spin relaxation rate $\Gamma_s$ for the heavy quark, based on pQCD to leading-log order of coupling constant $g$.
We have formulated three different methods to evaluate $\Gamma_s$: 
1) the Green-Kubo formula based on the source-source correlator in the spin hydrodynamic regime, 2) the spin density correlator in the strict hydrodynamic regime, and 3) the quantum kinetic equation for the heavy-quark spin distribution. 
While each method demonstrates a different view on spin dynamics,
all of these lead to the same result $\Gamma_s\sim g^4\log(1/g)T (T/M)^2$ [see eqs.~\eqref{eq:Gamma-spin}, \eqref{eq:Gamma-spin2}, and \eqref{eq:Gamma-spin3}].
Thanks to the additional $(T/M)^2$ factor, the heavy quark spin shows a parametrically slow dynamics compared to other non-hydrodynamic modes. 
This scale hierarchy $\Gamma_s \ll \Gamma$, where $\Gamma$ is the relaxation rate for other non-hydrodynamic modes, guarantees the existence of the spin hydrodynamic regime as a well-defined example of Hydro+~\cite{Stephanov:2017ghc}.

Several outlooks related to the present paper are in order.
While we rely on finite-temperature pQCD to evaluate the spin relaxation rate, 
our formulation --- in particular the heavy quark rotational viscosity, $\etas$ in terms of the source or spin correlation functions --- is applicable even in the strong-coupling regime, where we have the AdS/CFT correspondence as another theoretical  tool~\cite{Mcinnes:2018wzw,Garbiso:2020puw,Chen:2020ath,Cartwright:2021qpp}, and in particular the methods first developed for holographic spin liquids in ref.~\cite{Gallegos:2020otk}.
It would be interesting to ask how we can formulate heavy-quark spin relaxation in the strong-coupling regime of QCD-like theories, such as $\Ncal = 4$ super Yang-Mills theory, in a manner similar to the heavy quark drag force~\cite{Herzog:2006gh,Gubser:2006bz,Casalderrey-Solana:2006fio}. 
It will give an important benchmark in another extreme limit of the theory, which would be useful in the phenomenological analysis of spin dynamics within the QGP created in heavy-ion collisions.
 
Another interesting outlook is to apply our result in astrophysics or condensed matter systems.
In fact, while we focus on the heavy-quark spin relaxation in QCD plasma in the present paper, our formulation works as well in other systems where spin is approximately conserved.
For example, the relaxation rate of proton spin in QED plasma can be studied by our methods with a minor modification.
Also, it is interesting to apply the present formulation to various spin systems in condensed matter physics, which deviate from the Heisenberg model only by a small symmetry-breaking perturbation. 
In this case, the effective field theory with symmetry breaking terms~(see, e.g., refs.~\cite{Gongyo:2016dzp,Hongo:2020xaw}) may be a useful starting point to describe the relaxation rate of spin in condensed matter physics. 
We leave all these as future work.

\acknowledgments
We thank Yoshimasa Hidaka for useful comments on the Kadanoff-Baym formalism. 
This work is supported by the U.S. Department of Energy, Office of Science, Office of Nuclear Physics under Award Number DE-FG0201ER41195, and within the framework of the Beam Energy Scan Theory (BEST) Topical Collaboration, and also partially by RIKEN iTHEMS Program (in particular, iTHEMS Non-Equilibrium Working Group and Mathematical Physics Working Group). 
X.~G.~H is supported by NSFC under Grant No.~12075061 and Shanghai NSF under Grant No.~20ZR1404100. M.~K. is supported, in part, by the U.S. Department of Energy grant DE-SC0012447.


\appendix

\section{Derivation of quantum kinetic equation from Kadanoff-Baym formalism}
\label{sec:Kadanoff-Baym}

In this appendix, we derive the quantum kinetic equation~\eqref{eq:eom-density-operator}-\eqref{eq:damping-term-density-operator} for heavy quarks based on the real-time Kadanoff-Baym formalism~\cite{Kadanoff-Baym1961}. 

Our starting point is the Schwinger-Dyson equation for heavy-quark Green's function in the real-time formalism, diagrammatically given by
\begin{equation}
 \begin{split}
  \parbox{2.5cm}{\vspace{6pt}\includegraphics[width=2.5cm]{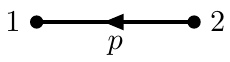}}
  &= 
  \parbox{2.5cm}{\vspace{6pt}\includegraphics[width=2.5cm]{fig-S12.pdf}}
  ~+~ 
  \sum_{a,b=1}^2~
  \parbox{4.0cm}{\vspace{0pt}\includegraphics[width=4.0cm]{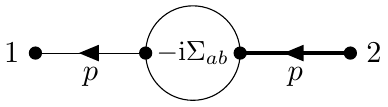}}
  \\
  &=
  \parbox{2.5cm}{\vspace{6pt}\includegraphics[width=2.5cm]{fig-S12.pdf}}
  ~+~
  \sum_{a,b=1}^2 ~
  \parbox{4.0cm}{\vspace{0pt}\includegraphics[width=4.0cm]{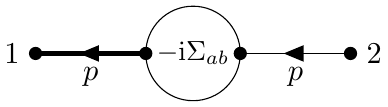}}\;\;,
 \end{split}
 \label{eq:Schwinger-Dyson-Diagram}
\end{equation}
where $\Sigma_{ab}~(a,b=1,2)$ denotes the heavy-quark self-energy in the $12$ basis on the Schwinger-Keldysh contour.
Note that Green's functions in the $ra$ basis are related to those in the $12$ basis as 
\begin{equation}
 \begin{split}
  S_{ra}
  &= \frac{1}{2} 
  \left( 
  S_{11} - S_{12} + S_{21} - S_{22}
  \right),
  \\
  S_{ar}
  &= \frac{1}{2} 
  \left( 
  S_{11} + S_{12} - S_{21} - S_{22}
  \right).
 \end{split}
\end{equation}
After a little cumbersome computation, 
we can rewrite the Schwinger-Dyson equation depicted in eq.~\eqref{eq:Schwinger-Dyson-Diagram} as follows:
\begin{equation}
 \begin{split}
  S_{12}
  &= S^{(0)}_{12}
  - \rmi S^{(0)}_{ra} \Sigma_{ra} S_{12}
  - \rmi S^{(0)}_{ra} \Sigma_{12} S_{ar}
  - \rmi S^{(0)}_{12} \Sigma_{ar} S_{ar}
  \\
  &= S^{(0)}_{12}
  - \rmi S_{ra} \Sigma_{ra} S^{(0)}_{12}
  - \rmi S_{ra} \Sigma_{12} S^{(0)}_{ar}
  - \rmi S_{12} \Sigma_{ar} S^{(0)}_{ar}.
 \end{split}
 \label{eq:Schwinger-Dyson1}
\end{equation}
Acting the free inverse propagator 
$(S^{(0)})^{-1}$ satisfying
$(S^{(0)})^{-1} S^{(0)}_{ra} = \rmi = S^{(0)}_{ar}  (S^{(0)})^{-1} $ 
and 
$(S^{(0)})^{-1} S^{(0)}_{12} 
 = 0 = S^{(0)}_{12} (S^{(0)})^{-1}$
to this equation, we obtain
the left and right Schwinger-Dyson equations
\begin{subequations}
 \label{eq:Schwinger-Dyson2}
  \begin{align}
  (S^{(0)})^{-1} S_{12}
  &= \Sigma_{ra} S_{12} + \Sigma_{12} S_{ar},
  \\
  S_{12} (S^{(0)})^{-1}
  &= S_{ra} \Sigma_{12} + S_{12} \Sigma_{ar} .
  \end{align}
\end{subequations}

We then take the difference of these two equations. 
Assuming the two-point function $S_{12}$ has a weak dependence on the center-of-mass coordinate, we can rely on the derivative expansion.
Moreover, recalling the explicit form of the free propagator
$(S^{(0)})^{-1} = \rmi \partial_t +\frac{1}{2M} \bnab^2$, we find that the difference of the Schwinger-Dyson equations becomes
\begin{equation}
 \begin{split}
  \left( 
  \partial_t + \frac{1}{M} \bp \cdot \bnab_x
  \right)  
  S_{12} (x,p)
  =& - \rmi
  \big[
  \Sigma_{ra} (x,p) \star S_{12} (x,p)
  + \Sigma_{12} (x,p) \star S_{ar} (x,p)
  \\
  &\hspace{14pt}
  - S_{ra} (x,p) \star \Sigma_{12} (x,p)
  - S_{12} (x,p) \star \Sigma_{ar} (x,p)
  \big] .
 \end{split}
 \label{eq:Bol0}
\end{equation}
where we introduced the Wigner transform 
of two-point functions 
and the Moyal product of $A (x,p)$ and $B(x,p)$ as
\begin{align}
  G (x,p) 
  &\equiv
  \int \frac{\diff^4 r}{(2\pi)^4}
  \rme^{- \rmi p^0 r^0 + \rmi \bp \cdot \br}
  G \left(
  x + \frac{1}{2} r , x - \frac{1}{2} r
  \right),
  \\
  A(x,p) \star B (x,p)
  &\equiv 
  \rme^{\frac{\rmi}{2} 
  (\partial_{x} \cdot \partial_{p^{\prime}} 
  - \partial_{x^{\prime}} \cdot \partial_{p})} 
  A(x,p) B (x^{\prime},p^{\prime})
  \Big|_{x^{\prime} = x, p^{\prime} = p} .
\end{align}

Expressing some parts of 
$S_{ra}, S_{ar}, \Sigma_{ra}$, and $\Sigma_{ar}$
in the $12$ basis, and reorganizing them,
we obtain a quantum transport equation for the real-time Green's function $S_{12}$ as 
\begin{equation}
 \begin{split}
  \left( 
  \partial_t + \frac{1}{M} \bp \cdot \bnab_x
  \right)
  S_{12} 
  + [ \rmi \Re \Sigma_{ra}, S_{12} ]_{\star}
  - [\Sigma_{12}, \im S_{ra}]_{\star}
  = 
  - \frac{\rmi}{2}
  \Big[ 
  \{ \Sigma_{21}, S_{12} \}_{\star}
  - \{ \Sigma_{12}, S_{21} \}_{\star}
  \Big],
 \end{split}
 \label{eq:Boltzmann}
\end{equation}
where we introduced a commutator 
$[A,B]_{\star} \equiv A \star B - B\star A$ and 
anti-commutator $\{A,B\}_{\star} \equiv A \star B + B\star A$ with the Moyal product, respectively.
For our purpose, we can neglect all $\bx$-dependence and the terms with commutators in the left-hand side since the collision term is captured by the right-hand side of this equation.
Moreover, expanding the Moyal product simplifies the right-hand side as $\{A,B\}_{\star} \simeq A B + BA$.
Then, by evaluating the right-hand side (the collision term) with a usual product, we identify this equation as the quantum kinetic equation for heavy quarks including spin, which agrees with eq.~\eqref{eq:eom-density-operator} in the main text.

Let us evaluate the self-energy appearing in the right-hand side of eq.~\eqref{eq:Boltzmann}.
As usual, the leading diagram is given by
\vspace{20pt}
\begin{equation}
 \begin{split}
 - \rmi \Sigma_{ab} (p)
  &=~
  \parbox{3.2cm}{\vspace{-22pt}\includegraphics[width=3.2cm]{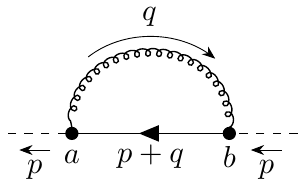}}  
  = (\rmi g)^2 C_2 (R)
  \int \frac{\diff^4 q}{(2\pi)^4} 
  D^{\beta\alpha}_{ba} (q)   
  H_\alpha (-\bq) S_{ab} (p+q) H_\beta (\bq),
 \end{split}
\end{equation}
where the interaction vertex $H_\alpha (\bq)$  between the heavy quark and gluon is given by eq.~\eqref{eq:heavy-quark-gluon-vertex}.
Relying on the quasi-particle approximation for heavy quarks, we parametrize the lesser and greater Green's functions as 
\begin{equation}
 S_{12} (t,\bp) = - \rho(t,\bp) 2 \pi \delta (p^0- E_{\bp})
 \quad \mathrm{and} \quad
 S_{21} (t,\bp) = [ \bm{1}_{2\times 2} - \rho (t,\bp) ] 2 \pi \delta (p^0- E_{\bp}),
\end{equation}
with a $2 \times 2$ distribution function $\rho (t,\bp)$.
As a result, we obtain the following expression for the self-energy:
\begin{equation}
 \begin{split}
  - \rmi \Sigma_{12} (p)
  &= g^2 C_2 (R)
  \int \frac{\diff^4 q}{(2\pi)^4} 
  H_\alpha (-\bq) 
  \rho (t,\bp+\bq)
  H_\beta (\bq)
  D^{\alpha\beta}_{21} (q)   
  2\pi \delta (p^0 + q^0 - E_{\bp + \bq})
  ,
  \\
  - \rmi \Sigma_{21} (p)
  &= g^2 C_2 (R)
  \int \frac{\diff^4 q}{(2\pi)^4} 
  H_\alpha (-\bq) 
  \big[ \rho (t,\bp+\bq) - \bm{1}_{2 \times 2} \big]
  H_\beta (\bq)
  D^{\alpha\beta}_{12} (q)   
  2\pi \delta (p^0 + q^0 - E_{\bp + \bq}).
 \end{split}
\end{equation}
Thanks to the quasi-particle approximation, 
the heavy-quark propagator is accompanied by the delta function, which we can eliminate by performing the $p^0$-integration.
Furthermore, we consider the dilute system of heavy quarks, and neglect $\Ocal(\rho^2)$-terms.
As a consequence, we derive a simplified form for the quantum kinetic equation as 
\begin{equation}
 \begin{split}
  \partial_t \rho 
  &\simeq 
  g^2 C_2 (R)
  \int \frac{\diff^4 q}{(2\pi)^4}
  \bigg[ 
  H_\alpha (\bq) 
  \rho (t,\bp-\bq)
  H_\beta (-\bq)
  D^{\alpha\beta}_{12} (q)   
  2\pi \delta ( q^0 + E_{\bp - \bq} - E_{\bp} )
  \\
  &\hspace{74pt}
  - \frac{1}{2} \tr 
  \big[
  H_\alpha (\bq) H_\beta (-\bq) 
  \big]
  D^{\alpha\beta}_{21} (q)   
  2\pi \delta (q^0 + E_{\bp - \bq} - E_{\bp})  
  \rho (t,\bp)
  \bigg],
 \end{split}
 \label{eq:quantum-kinetic-spin}
\end{equation}
where we have changed the integration variable as $q \to -q$ and used the identity $D_{12}^{\alpha\beta} (-q) = D_{21}^{\alpha\beta} (q)$.
Besides, we also used the following simplification 
for the matrix structure in the spin space after $q$ integration (see the Appendix of ref.~\cite{Li:2019qkf}): 
\begin{equation}
 \big[
  H_\alpha (\bq) H_\beta (-\bq) \rho (t,\bp)
  + \rho (t,\bp) H_\alpha (\bq) H_\beta (-\bq)
  \big]D^{\alpha\beta}_{21} (q)
 \to 
 \tr 
 \big[
 H_\alpha (\bq) H_\beta (-\bq) 
 \big]D^{\alpha\beta}_{21} (q)
 \rho (t,\bp).
\end{equation}
One sees that eq.~\eqref{eq:quantum-kinetic-spin} indeed coincides with the eq.~\eqref{eq:eom-density-operator} in the main text.

\bibliographystyle{utphys}
\bibliography{spin-damping}

\providecommand{\href}[2]{#2}\begingroup\raggedright\begin{thebibliography}{10}

\bibitem{STAR:2017ckg}
{\bfseries STAR} Collaboration, L.~Adamczyk {\em et~al.}, ``{Global $\Lambda$
  hyperon polarization in nuclear collisions: evidence for the most vortical
  fluid},'' \href{http://dx.doi.org/10.1038/nature23004}{{\em Nature}
  {\bfseries 548} (2017) 62--65},
\href{http://arxiv.org/abs/1701.06657}{{\ttfamily arXiv:1701.06657 [nucl-ex]}}.

\bibitem{STAR:2019erd}
{\bfseries STAR} Collaboration, J.~Adam {\em et~al.}, ``{Polarization of
  $\Lambda$ ($\bar{\Lambda}$) hyperons along the beam direction in Au+Au
  collisions at $\sqrt{s_{_{NN}}}$ = 200 GeV},''
  \href{http://dx.doi.org/10.1103/PhysRevLett.123.132301}{{\em Phys. Rev.
  Lett.} {\bfseries 123} no.~13, (2019) 132301},
  \href{http://arxiv.org/abs/1905.11917}{{\ttfamily arXiv:1905.11917
  [nucl-ex]}}.

\bibitem{ALICE:2019aid}
{\bfseries ALICE} Collaboration, S.~Acharya {\em et~al.}, ``{Evidence of
  Spin-Orbital Angular Momentum Interactions in Relativistic Heavy-Ion
  Collisions},'' \href{http://dx.doi.org/10.1103/PhysRevLett.125.012301}{{\em
  Phys. Rev. Lett.} {\bfseries 125} no.~1, (2020) 012301},
  \href{http://arxiv.org/abs/1910.14408}{{\ttfamily arXiv:1910.14408
  [nucl-ex]}}.

\bibitem{STAR:2020xbm}
{\bfseries STAR} Collaboration, J.~Adam {\em et~al.}, ``{Global Polarization of
  $\Xi$ and $\Omega$ Hyperons in Au+Au Collisions at $\sqrt {s_{NN}}$ = 200
  GeV},'' \href{http://dx.doi.org/10.1103/PhysRevLett.126.162301}{{\em Phys.
  Rev. Lett.} {\bfseries 126} no.~16, (2021) 162301},
  \href{http://arxiv.org/abs/2012.13601}{{\ttfamily arXiv:2012.13601
  [nucl-ex]}}.

\bibitem{Gao:2019znl}
J.-H. Gao and Z.-T. Liang, ``{Relativistic Quantum Kinetic Theory for Massive
  Fermions and Spin Effects},''
  \href{http://dx.doi.org/10.1103/PhysRevD.100.056021}{{\em Phys. Rev. D}
  {\bfseries 100} no.~5, (2019) 056021},
  \href{http://arxiv.org/abs/1902.06510}{{\ttfamily arXiv:1902.06510
  [hep-ph]}}.

\bibitem{Weickgenannt:2019dks}
N.~Weickgenannt, X.-L. Sheng, E.~Speranza, Q.~Wang, and D.~H. Rischke,
  ``{Kinetic theory for massive spin-1/2 particles from the Wigner-function
  formalism},'' \href{http://dx.doi.org/10.1103/PhysRevD.100.056018}{{\em Phys.
  Rev. D} {\bfseries 100} no.~5, (2019) 056018},
  \href{http://arxiv.org/abs/1902.06513}{{\ttfamily arXiv:1902.06513
  [hep-ph]}}.

\bibitem{Hattori:2019ahi}
K.~Hattori, Y.~Hidaka, and D.-L. Yang, ``{Axial Kinetic Theory and Spin
  Transport for Fermions with Arbitrary Mass},''
  \href{http://dx.doi.org/10.1103/PhysRevD.100.096011}{{\em Phys. Rev. D}
  {\bfseries 100} no.~9, (2019) 096011},
  \href{http://arxiv.org/abs/1903.01653}{{\ttfamily arXiv:1903.01653
  [hep-ph]}}.

\bibitem{Wang:2019moi}
Z.~Wang, X.~Guo, S.~Shi, and P.~Zhuang, ``{Mass Correction to Chiral Kinetic
  Equations},'' \href{http://dx.doi.org/10.1103/PhysRevD.100.014015}{{\em Phys.
  Rev. D} {\bfseries 100} no.~1, (2019) 014015},
  \href{http://arxiv.org/abs/1903.03461}{{\ttfamily arXiv:1903.03461
  [hep-ph]}}.

\bibitem{Li:2019qkf}
S.~Li and H.-U. Yee, ``{Quantum Kinetic Theory of Spin Polarization of Massive
  Quarks in Perturbative QCD: Leading Log},''
  \href{http://dx.doi.org/10.1103/PhysRevD.100.056022}{{\em Phys. Rev. D}
  {\bfseries 100} no.~5, (2019) 056022},
  \href{http://arxiv.org/abs/1905.10463}{{\ttfamily arXiv:1905.10463
  [hep-ph]}}.

\bibitem{Kapusta:2019sad}
J.~I. Kapusta, E.~Rrapaj, and S.~Rudaz, ``{Relaxation Time for Strange Quark
  Spin in Rotating Quark-Gluon Plasma},''
  \href{http://dx.doi.org/10.1103/PhysRevC.101.024907}{{\em Phys. Rev. C}
  {\bfseries 101} no.~2, (2020) 024907},
  \href{http://arxiv.org/abs/1907.10750}{{\ttfamily arXiv:1907.10750
  [nucl-th]}}.

\bibitem{Liu:2019krs}
S.~Y.~F. Liu, Y.~Sun, and C.~M. Ko, ``{Spin Polarizations in a Covariant
  Angular-Momentum-Conserved Chiral Transport Model},''
  \href{http://dx.doi.org/10.1103/PhysRevLett.125.062301}{{\em Phys. Rev.
  Lett.} {\bfseries 125} no.~6, (2020) 062301},
  \href{http://arxiv.org/abs/1910.06774}{{\ttfamily arXiv:1910.06774
  [nucl-th]}}.

\bibitem{Yang:2020hri}
D.-L. Yang, K.~Hattori, and Y.~Hidaka, ``{Effective quantum kinetic theory for
  spin transport of fermions with collsional effects},''
  \href{http://dx.doi.org/10.1007/JHEP07(2020)070}{{\em JHEP} {\bfseries 07}
  (2020) 070}, \href{http://arxiv.org/abs/2002.02612}{{\ttfamily
  arXiv:2002.02612 [hep-ph]}}.

\bibitem{Liu:2020flb}
Y.-C. Liu, K.~Mameda, and X.-G. Huang, ``{Covariant Spin Kinetic Theory I:
  Collisionless Limit},''
  \href{http://dx.doi.org/10.1088/1674-1137/44/9/094101}{{\em Chin. Phys. C}
  {\bfseries 44} no.~9, (2020) 094101},
  \href{http://arxiv.org/abs/2002.03753}{{\ttfamily arXiv:2002.03753
  [hep-ph]}}.

\bibitem{Weickgenannt:2020aaf}
N.~Weickgenannt, E.~Speranza, X.-l. Sheng, Q.~Wang, and D.~H. Rischke,
  ``{Generating Spin Polarization from Vorticity through Nonlocal
  Collisions},'' \href{http://dx.doi.org/10.1103/PhysRevLett.127.052301}{{\em
  Phys. Rev. Lett.} {\bfseries 127} no.~5, (2021) 052301},
  \href{http://arxiv.org/abs/2005.01506}{{\ttfamily arXiv:2005.01506
  [hep-ph]}}.

\bibitem{Weickgenannt:2021cuo}
N.~Weickgenannt, E.~Speranza, X.-l. Sheng, Q.~Wang, and D.~H. Rischke,
  ``{Derivation of the nonlocal collision term in the relativistic Boltzmann
  equation for massive spin-1/2 particles from quantum field theory},''
  \href{http://dx.doi.org/10.1103/PhysRevD.104.016022}{{\em Phys. Rev. D}
  {\bfseries 104} no.~1, (2021) 016022},
  \href{http://arxiv.org/abs/2103.04896}{{\ttfamily arXiv:2103.04896
  [nucl-th]}}.

\bibitem{Sheng:2021kfc}
X.-L. Sheng, N.~Weickgenannt, E.~Speranza, D.~H. Rischke, and Q.~Wang, ``{From
  Kadanoff-Baym to Boltzmann equations for massive spin-1/2 fermions},''
  \href{http://dx.doi.org/10.1103/PhysRevD.104.016029}{{\em Phys. Rev. D}
  {\bfseries 104} no.~1, (2021) 016029},
  \href{http://arxiv.org/abs/2103.10636}{{\ttfamily arXiv:2103.10636
  [nucl-th]}}.

\bibitem{Lin:2021mvw}
S.~Lin, ``{Quantum Kinetic Theory for Quantum Electrodynamics},''
  \href{http://arxiv.org/abs/2109.00184}{{\ttfamily arXiv:2109.00184
  [hep-ph]}}.

\bibitem{Florkowski:2017ruc}
W.~Florkowski, B.~Friman, A.~Jaiswal, and E.~Speranza, ``{Relativistic fluid
  dynamics with spin},''
  \href{http://dx.doi.org/10.1103/PhysRevC.97.041901}{{\em Phys. Rev. C}
  {\bfseries 97} no.~4, (2018) 041901},
  \href{http://arxiv.org/abs/1705.00587}{{\ttfamily arXiv:1705.00587
  [nucl-th]}}.

\bibitem{Florkowski:2018fap}
W.~Florkowski, A.~Kumar, and R.~Ryblewski, ``{Relativistic hydrodynamics for
  spin-polarized fluids},''
  \href{http://dx.doi.org/10.1016/j.ppnp.2019.07.001}{{\em Prog. Part. Nucl.
  Phys.} {\bfseries 108} (2019) 103709},
  \href{http://arxiv.org/abs/1811.04409}{{\ttfamily arXiv:1811.04409
  [nucl-th]}}.

\bibitem{Hattori:2019lfp}
K.~Hattori, M.~Hongo, X.-G. Huang, M.~Matsuo, and H.~Taya, ``{Fate of spin
  polarization in a relativistic fluid: An entropy-current analysis},''
  \href{http://dx.doi.org/10.1016/j.physletb.2019.05.040}{{\em Phys. Lett. B}
  {\bfseries 795} (2019) 100--106},
  \href{http://arxiv.org/abs/1901.06615}{{\ttfamily arXiv:1901.06615
  [hep-th]}}.

\bibitem{Fukushima:2020ucl}
K.~Fukushima and S.~Pu, ``{Spin hydrodynamics and symmetric energy-momentum
  tensors \textendash{} A current induced by the spin vorticity
  \textendash{}},''
  \href{http://dx.doi.org/10.1016/j.physletb.2021.136346}{{\em Phys. Lett. B}
  {\bfseries 817} (2021) 136346},
  \href{http://arxiv.org/abs/2010.01608}{{\ttfamily arXiv:2010.01608
  [hep-th]}}.

\bibitem{Bhadury:2020puc}
S.~Bhadury, W.~Florkowski, A.~Jaiswal, A.~Kumar, and R.~Ryblewski,
  ``{Relativistic dissipative spin dynamics in the relaxation time
  approximation},''
  \href{http://dx.doi.org/10.1016/j.physletb.2021.136096}{{\em Phys. Lett. B}
  {\bfseries 814} (2021) 136096},
  \href{http://arxiv.org/abs/2002.03937}{{\ttfamily arXiv:2002.03937
  [hep-ph]}}.

\bibitem{Shi:2020htn}
S.~Shi, C.~Gale, and S.~Jeon, ``{From chiral kinetic theory to relativistic
  viscous spin hydrodynamics},''
  \href{http://dx.doi.org/10.1103/PhysRevC.103.044906}{{\em Phys. Rev. C}
  {\bfseries 103} no.~4, (2021) 044906},
  \href{http://arxiv.org/abs/2008.08618}{{\ttfamily arXiv:2008.08618
  [nucl-th]}}.

\bibitem{Li:2020eon}
S.~Li, M.~A. Stephanov, and H.-U. Yee, ``{Nondissipative Second-Order
  Transport, Spin, and Pseudogauge Transformations in Hydrodynamics},''
  \href{http://dx.doi.org/10.1103/PhysRevLett.127.082302}{{\em Phys. Rev.
  Lett.} {\bfseries 127} no.~8, (2021) 082302},
  \href{http://arxiv.org/abs/2011.12318}{{\ttfamily arXiv:2011.12318
  [hep-th]}}.

\bibitem{Gallegos:2021bzp}
A.~D. Gallegos, U.~G\"ursoy, and A.~Yarom, ``{Hydrodynamics of spin
  currents},'' \href{http://dx.doi.org/10.21468/SciPostPhys.11.2.041}{{\em
  SciPost Phys.} {\bfseries 11} (2021) 041},
  \href{http://arxiv.org/abs/2101.04759}{{\ttfamily arXiv:2101.04759
  [hep-th]}}.

\bibitem{Liu:2021uhn}
S.~Y.~F. Liu and Y.~Yin, ``{Spin polarization induced by the hydrodynamic
  gradients},'' \href{http://dx.doi.org/10.1007/JHEP07(2021)188}{{\em JHEP}
  {\bfseries 07} (2021) 188}, \href{http://arxiv.org/abs/2103.09200}{{\ttfamily
  arXiv:2103.09200 [hep-ph]}}.

\bibitem{She:2021lhe}
D.~She, A.~Huang, D.~Hou, and J.~Liao, ``{Relativistic Viscous Hydrodynamics
  with Angular Momentum},'' \href{http://arxiv.org/abs/2105.04060}{{\ttfamily
  arXiv:2105.04060 [nucl-th]}}.

\bibitem{Hongo:2021ona}
M.~Hongo, X.-G. Huang, M.~Kaminski, M.~Stephanov, and H.-U. Yee,
  ``{Relativistic spin hydrodynamics with torsion and linear response theory
  for spin relaxation},'' \href{http://dx.doi.org/10.1007/JHEP11(2021)150}{{\em
  JHEP} {\bfseries 11} (2021) 150},
  \href{http://arxiv.org/abs/2107.14231}{{\ttfamily arXiv:2107.14231
  [hep-th]}}.

\bibitem{Peng:2021ago}
H.-H. Peng, J.-J. Zhang, X.-L. Sheng, and Q.~Wang, ``{Ideal Spin Hydrodynamics
  from the Wigner Function Approach},''
  \href{http://dx.doi.org/10.1088/0256-307X/38/11/116701}{{\em Chin. Phys.
  Lett.} {\bfseries 38} no.~11, (2021) 116701},
  \href{http://arxiv.org/abs/2107.00448}{{\ttfamily arXiv:2107.00448
  [hep-th]}}.

\bibitem{Becattini:2007nd}
F.~Becattini and F.~Piccinini, ``{The Ideal relativistic spinning gas:
  Polarization and spectra},''
  \href{http://dx.doi.org/10.1016/j.aop.2008.01.001}{{\em Annals Phys.}
  {\bfseries 323} (2008) 2452--2473},
  \href{http://arxiv.org/abs/0710.5694}{{\ttfamily arXiv:0710.5694 [nucl-th]}}.

\bibitem{Becattini:2009wh}
F.~Becattini and L.~Tinti, ``{The Ideal relativistic rotating gas as a perfect
  fluid with spin},'' \href{http://dx.doi.org/10.1016/j.aop.2010.03.007}{{\em
  Annals Phys.} {\bfseries 325} (2010) 1566--1594},
  \href{http://arxiv.org/abs/0911.0864}{{\ttfamily arXiv:0911.0864 [gr-qc]}}.

\bibitem{Becattini:2014yxa}
F.~Becattini, L.~Bucciantini, E.~Grossi, and L.~Tinti, ``{Local thermodynamical
  equilibrium and the beta frame for a quantum relativistic fluid},''
  \href{http://dx.doi.org/10.1140/epjc/s10052-015-3384-y}{{\em Eur. Phys. J. C}
  {\bfseries 75} no.~5, (2015) 191},
  \href{http://arxiv.org/abs/1403.6265}{{\ttfamily arXiv:1403.6265 [hep-th]}}.

\bibitem{Palermo:2021hlf}
A.~Palermo, M.~Buzzegoli, and F.~Becattini, ``{Exact equilibrium distributions
  in statistical quantum field theory with rotation and acceleration: Dirac
  field},'' \href{http://dx.doi.org/10.1007/JHEP10(2021)077}{{\em JHEP}
  {\bfseries 10} (2021) 077}, \href{http://arxiv.org/abs/2106.08340}{{\ttfamily
  arXiv:2106.08340 [hep-th]}}.

\bibitem{Becattini:2012pp}
F.~Becattini and L.~Tinti, ``{Nonequilibrium Thermodynamical Inequivalence of
  Quantum Stress-energy and Spin Tensors},''
  \href{http://dx.doi.org/10.1103/PhysRevD.87.025029}{{\em Phys. Rev. D}
  {\bfseries 87} no.~2, (2013) 025029},
  \href{http://arxiv.org/abs/1209.6212}{{\ttfamily arXiv:1209.6212 [hep-th]}}.

\bibitem{Becattini:2018duy}
F.~Becattini, W.~Florkowski, and E.~Speranza, ``{Spin tensor and its role in
  non-equilibrium thermodynamics},''
  \href{http://dx.doi.org/10.1016/j.physletb.2018.12.016}{{\em Phys. Lett. B}
  {\bfseries 789} (2019) 419--425},
  \href{http://arxiv.org/abs/1807.10994}{{\ttfamily arXiv:1807.10994
  [hep-th]}}.

\bibitem{Speranza:2020ilk}
E.~Speranza and N.~Weickgenannt, ``{Spin tensor and pseudo-gauges: from nuclear
  collisions to gravitational physics},''
  \href{http://dx.doi.org/10.1140/epja/s10050-021-00455-2}{{\em Eur. Phys. J.
  A} {\bfseries 57} no.~5, (2021) 155},
  \href{http://arxiv.org/abs/2007.00138}{{\ttfamily arXiv:2007.00138
  [nucl-th]}}.

\bibitem{Braaten:1989mz}
E.~Braaten and R.~D. Pisarski, ``{Soft Amplitudes in Hot Gauge Theories: A
  General Analysis},''
  \href{http://dx.doi.org/10.1016/0550-3213(90)90508-B}{{\em Nucl. Phys. B}
  {\bfseries 337} (1990) 569--634}.

\bibitem{LeBellac2000}
M.~Le~Bellac, {\em Thermal field theory}.
\newblock Cambridge University Press, 2000.

\bibitem{Blaizot:2001nr}
J.-P. Blaizot and E.~Iancu, ``{The Quark gluon plasma: Collective dynamics and
  hard thermal loops},''
  \href{http://dx.doi.org/10.1016/S0370-1573(01)00061-8}{{\em Phys. Rept.}
  {\bfseries 359} (2002) 355--528},
  \href{http://arxiv.org/abs/hep-ph/0101103}{{\ttfamily arXiv:hep-ph/0101103}}.

\bibitem{Stephanov:2017ghc}
M.~Stephanov and Y.~Yin, ``{Hydrodynamics with parametric slowing down and
  fluctuations near the critical point},''
  \href{http://dx.doi.org/10.1103/PhysRevD.98.036006}{{\em Phys. Rev. D}
  {\bfseries 98} no.~3, (2018) 036006},
  \href{http://arxiv.org/abs/1712.10305}{{\ttfamily arXiv:1712.10305
  [nucl-th]}}.

\bibitem{Manohar-Wise2000}
A.~V. Manohar and M.~B. Wise, {\em Heavy Quark Physics}.
\newblock Cambridge Monographs on Particle Physics, Nuclear Physics and
  Cosmology. Cambridge University Press, 2000.

\bibitem{Hidaka:2022dmn}
Y.~Hidaka, S.~Pu, Q.~Wang, and D.-L. Yang, ``{Foundations and Applications of
  Quantum Kinetic Theory},'' \href{http://arxiv.org/abs/2201.07644}{{\ttfamily
  arXiv:2201.07644 [hep-ph]}}.

\bibitem{Kadanoff-Martin1963}
L.~P. Kadanoff and P.~C. Martin, ``Hydrodynamic equations and correlation
  functions,''
  \href{http://dx.doi.org/https://doi.org/10.1016/0003-4916(63)90078-2}{{\em
  Annals of Physics} {\bfseries 24} (1963) 419 -- 469}.

\bibitem{Jeon:1994if}
S.~Jeon, ``{Hydrodynamic transport coefficients in relativistic scalar field
  theory},'' \href{http://dx.doi.org/10.1103/PhysRevD.52.3591}{{\em Phys. Rev.
  D} {\bfseries 52} (1995) 3591--3642},
  \href{http://arxiv.org/abs/hep-ph/9409250}{{\ttfamily arXiv:hep-ph/9409250}}.

\bibitem{ValleBasagoiti:2002ir}
M.~A. Valle~Basagoiti, ``{Transport coefficients and ladder summation in hot
  gauge theories},'' \href{http://dx.doi.org/10.1103/PhysRevD.66.045005}{{\em
  Phys. Rev. D} {\bfseries 66} (2002) 045005},
  \href{http://arxiv.org/abs/hep-ph/0204334}{{\ttfamily arXiv:hep-ph/0204334}}.

\bibitem{Jimenez-Alba:2015bia}
A.~Jimenez-Alba and H.-U. Yee, ``{Second order transport coefficient from the
  chiral anomaly at weak coupling: Diagrammatic resummation},''
  \href{http://dx.doi.org/10.1103/PhysRevD.92.014023}{{\em Phys. Rev. D}
  {\bfseries 92} no.~1, (2015) 014023},
  \href{http://arxiv.org/abs/1504.05866}{{\ttfamily arXiv:1504.05866
  [hep-ph]}}.

\bibitem{Aarts:2002tn}
G.~Aarts and J.~M. Martinez~Resco, ``{Ward identity and electrical conductivity
  in hot QED},'' \href{http://dx.doi.org/10.1088/1126-6708/2002/11/022}{{\em
  JHEP} {\bfseries 11} (2002) 022},
  \href{http://arxiv.org/abs/hep-ph/0209048}{{\ttfamily arXiv:hep-ph/0209048}}.

\bibitem{Hidaka:2010gh}
Y.~Hidaka and T.~Kunihiro, ``{Renormalized Linear Kinetic Theory as Derived
  from Quantum Field Theory: A Novel diagrammatic method for computing
  transport coefficients},''
  \href{http://dx.doi.org/10.1103/PhysRevD.83.076004}{{\em Phys. Rev. D}
  {\bfseries 83} (2011) 076004},
  \href{http://arxiv.org/abs/1009.5154}{{\ttfamily arXiv:1009.5154 [hep-ph]}}.

\bibitem{Moore:2004tg}
G.~D. Moore and D.~Teaney, ``{How much do heavy quarks thermalize in a heavy
  ion collision?},'' \href{http://dx.doi.org/10.1103/PhysRevC.71.064904}{{\em
  Phys. Rev. C} {\bfseries 71} (2005) 064904},
  \href{http://arxiv.org/abs/hep-ph/0412346}{{\ttfamily arXiv:hep-ph/0412346}}.

\bibitem{Baym:1990uj}
G.~Baym, H.~Monien, C.~J. Pethick, and D.~G. Ravenhall, ``{Transverse
  Interactions and Transport in Relativistic Quark - Gluon and Electromagnetic
  Plasmas},'' \href{http://dx.doi.org/10.1103/PhysRevLett.64.1867}{{\em Phys.
  Rev. Lett.} {\bfseries 64} (1990) 1867--1870}.

\bibitem{Arnold:2000dr}
P.~B. Arnold, G.~D. Moore, and L.~G. Yaffe, ``{Transport coefficients in high
  temperature gauge theories. 1. Leading log results},''
  \href{http://dx.doi.org/10.1088/1126-6708/2000/11/001}{{\em JHEP} {\bfseries
  11} (2000) 001}, \href{http://arxiv.org/abs/hep-ph/0010177}{{\ttfamily
  arXiv:hep-ph/0010177}}.

\bibitem{Schwinger:1960qe}
J.~S. Schwinger, ``{Brownian motion of a quantum oscillator},''
\href{http://dx.doi.org/10.1063/1.1703727}{{\em J. Math. Phys.} {\bfseries 2}
  (1961) 407--432}.

\bibitem{Keldysh:1964ud}
L.~V. Keldysh, ``{Diagram technique for nonequilibrium processes},'' {\em Zh.
  Eksp. Teor. Fiz.} {\bfseries 47} (1964) 1515--1527.

\bibitem{Landau:Kinetic}
E.~M. Lifshitz and L.~P. Pitaevskii, {\em Physical kinetics}.
\newblock Butterworth Heinemann, Oxford, UK, 1981.

\bibitem{Lindblad:1975ef}
G.~Lindblad, ``{On the Generators of Quantum Dynamical Semigroups},''
  \href{http://dx.doi.org/10.1007/BF01608499}{{\em Commun. Math. Phys.}
  {\bfseries 48} (1976) 119}.

\bibitem{Mcinnes:2018wzw}
B.~Mcinnes, ``{Applied holography of the AdS$_5$\textendash{}Kerr
  space\textendash{}time},''
  \href{http://dx.doi.org/10.1142/S0217751X19501380}{{\em Int. J. Mod. Phys. A}
  {\bfseries 34} no.~24, (2019) 1950138},
  \href{http://arxiv.org/abs/1803.02528}{{\ttfamily arXiv:1803.02528
  [hep-ph]}}.

\bibitem{Garbiso:2020puw}
M.~Garbiso and M.~Kaminski, ``{Hydrodynamics of simply spinning black holes \&
  hydrodynamics for spinning quantum fluids},''
  \href{http://dx.doi.org/10.1007/JHEP12(2020)112}{{\em JHEP} {\bfseries 12}
  (2020) 112}, \href{http://arxiv.org/abs/2007.04345}{{\ttfamily
  arXiv:2007.04345 [hep-th]}}.

\bibitem{Chen:2020ath}
X.~Chen, L.~Zhang, D.~Li, D.~Hou, and M.~Huang, ``{Gluodynamics and
  deconfinement phase transition under rotation from holography},''
  \href{http://dx.doi.org/10.1007/JHEP07(2021)132}{{\em JHEP} {\bfseries 07}
  (2021) 132}, \href{http://arxiv.org/abs/2010.14478}{{\ttfamily
  arXiv:2010.14478 [hep-ph]}}.

\bibitem{Cartwright:2021qpp}
C.~Cartwright, M.~G. Amano, M.~Kaminski, J.~Noronha, and E.~Speranza,
  ``{Convergence of hydrodynamics in rapidly spinning strongly coupled
  plasma},'' \href{http://arxiv.org/abs/2112.10781}{{\ttfamily arXiv:2112.10781
  [hep-th]}}.

\bibitem{Gallegos:2020otk}
A.~Gallegos and U.~G\"ursoy, ``{Holographic spin liquids and Lovelock
  Chern-Simons gravity},''
  \href{http://dx.doi.org/10.1007/JHEP11(2020)151}{{\em JHEP} {\bfseries 11}
  (2020) 151}, \href{http://arxiv.org/abs/2004.05148}{{\ttfamily
  arXiv:2004.05148 [hep-th]}}.

\bibitem{Herzog:2006gh}
C.~P. Herzog, A.~Karch, P.~Kovtun, C.~Kozcaz, and L.~G. Yaffe, ``{Energy loss
  of a heavy quark moving through N=4 supersymmetric Yang-Mills plasma},''
  \href{http://dx.doi.org/10.1088/1126-6708/2006/07/013}{{\em JHEP} {\bfseries
  07} (2006) 013}, \href{http://arxiv.org/abs/hep-th/0605158}{{\ttfamily
  arXiv:hep-th/0605158}}.

\bibitem{Gubser:2006bz}
S.~S. Gubser, ``{Drag force in AdS/CFT},''
  \href{http://dx.doi.org/10.1103/PhysRevD.74.126005}{{\em Phys. Rev. D}
  {\bfseries 74} (2006) 126005},
  \href{http://arxiv.org/abs/hep-th/0605182}{{\ttfamily arXiv:hep-th/0605182}}.

\bibitem{Casalderrey-Solana:2006fio}
J.~Casalderrey-Solana and D.~Teaney, ``{Heavy quark diffusion in strongly
  coupled N=4 Yang-Mills},''
  \href{http://dx.doi.org/10.1103/PhysRevD.74.085012}{{\em Phys. Rev. D}
  {\bfseries 74} (2006) 085012},
  \href{http://arxiv.org/abs/hep-ph/0605199}{{\ttfamily arXiv:hep-ph/0605199}}.

\bibitem{Gongyo:2016dzp}
S.~Gongyo, Y.~Kikuchi, T.~Hyodo, and T.~Kunihiro, ``{Effective field theory and
  the scattering process for magnons in ferromagnets, antiferromagnets, and
  ferrimagnets},'' \href{http://dx.doi.org/10.1093/ptep/ptw095}{{\em PTEP}
  {\bfseries 2016} no.~8, (2016) 083B01},
  \href{http://arxiv.org/abs/1602.08692}{{\ttfamily arXiv:1602.08692
  [cond-mat.str-el]}}.

\bibitem{Hongo:2020xaw}
M.~Hongo, T.~Fujimori, T.~Misumi, M.~Nitta, and N.~Sakai, ``{Effective field
  theory of magnons: Chiral magnets and the Schwinger mechanism},''
  \href{http://dx.doi.org/10.1103/PhysRevB.104.134403}{{\em Phys. Rev. B}
  {\bfseries 104} no.~13, (2021) 134403},
  \href{http://arxiv.org/abs/2009.06694}{{\ttfamily arXiv:2009.06694
  [cond-mat.mes-hall]}}.

\bibitem{Kadanoff-Baym1961}
G.~Baym and L.~P. Kadanoff, ``Conservation laws and correlation functions,''
  \href{http://dx.doi.org/10.1103/PhysRev.124.287}{{\em Phys. Rev.} {\bfseries
  124} (1961) 287--299}.

\end{thebibliography}\endgroup
\end{document}